\documentclass[reprint,superscriptaddress]{revtex4-2}

\usepackage{adjustbox}
\usepackage{amsmath}
\usepackage{amssymb}
\usepackage{amsthm}
\usepackage{bbm}
\usepackage{color}
\usepackage{float}
\usepackage[T1]{fontenc}
\usepackage{graphicx}
\usepackage{hyperref}
\usepackage[utf8]{inputenc}
\usepackage{physics}
\usepackage{times}
\usepackage{txfonts}
\usepackage{xcolor}
\usepackage{slashed}
\usepackage[normalem]{ulem}

\hypersetup{
  colorlinks=true,linkcolor=blue,citecolor=blue,
  filecolor=blue,urlcolor=blue,breaklinks=true
}

\newcommand{\orcid}[1]{\href{https://orcid.org/#1}
  {\includegraphics[width=7pt]{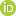}}}

\theoremstyle{definition}

\def\be{\begin{equation}}
        \def\ee{\end{equation}}

\def\bc{\begin{center}}
        \def\ec{\end{center}}
\def\bal{\begin{align}}
        \def\eal{\end{align}}

\begin{document}

\title{The Dirac equation: historical context, comparisons with the
  Schr\"{o}dinger and Klein-Gordon equations and elementary
  consequences.
}

\author{Thiago T. Tsutsui\orcid{0009-0001-1654-0330}}
\email{takajitsutsui@gmail.com}
\affiliation{
  Programa de Pós-Graduação em Ciências/Física,
  Universidade Estadual de Ponta Grossa,
  84030-900 Ponta Grossa, Paraná, Brazil
}

\author{Edilberto O. Silva\orcid{0000-0002-0297-5747}}
\email{edilberto.silva@ufma.br}
\affiliation{
  Departamento de F\'{i}sica,
  Universidade Federal do Maranh\~{a}o,
  Campus Universit\'{a}rio do Bacanga,
  65085-580 S\~{a}o Lu\'{i}, Maranhão , Brazil}

\author{Antonio S. M. de Castro\orcid{0000-0002-1521-9342}}
\email{asmcastro@uepg.br}
\affiliation{
  Programa de Pós-Graduação em Ciências/Física,
  Universidade Estadual de Ponta Grossa,
  84030-900 Ponta Grossa, Paraná, Brazil
}
\affiliation{
  Departamento de Física,
  Universidade Estadual de Ponta Grossa
  84030-900 Ponta Grossa, Paraná, Brazil
}

\author{Fabiano M. Andrade\orcid{0000-0001-5383-6168}}
\email{fmandrade@uepg.br}
\affiliation{
  Programa de Pós-Graduação em Ciências/Física,
  Universidade Estadual de Ponta Grossa,
  84030-900 Ponta Grossa, Paraná, Brazil
}
\affiliation{
  Departamento de Matemática e Estatística,
  Universidade Estadual de Ponta Grossa,
  84030-900 Ponta Grossa, Paraná, Brazil
}
\affiliation{
  Departamento de Física,
  Universidade Federal do Paraná,
  81531-980 Curitiba, Paraná, Brazil
}

\date{\today}

\begin{abstract}
This paper offers educational insight into the Dirac equation, examining its historical context and contrasting it with the earlier Schrödinger and Klein-Gordon equations. The comparison highlights their Lorentz transformation symmetry and potential probabilistic interpretations. We explicitly solve the free particle dynamics in Dirac's model, revealing the emergence of solutions with negative energy. In this discussion, we examine the Dirac Sea Hypothesis and explore the inherent helicity of the solutions. Additionally, we demonstrate how the Dirac equation accounts for spin and derive the Pauli equation in the non-relativistic limit. Through the Foldy-Wouthuysen transformation, we reveal how the equation incorporates spin-orbit interaction and other relativistic effects, ultimately leading to the hydrogen fine structure. A section on relativistic covariant notation is included to emphasize the invariance of the Dirac equation, along with more refined formulations of both the KG and Dirac equations. Designed for undergraduate students with an interest in the Dirac equation, this resource provides a historical perspective without being purely theoretical. Our approach underscores the significance of a pedagogical method that combines historical and comparative elements to profoundly understand the role of the Dirac equation in modern physics.
\\
\noindent \textbf{Eur. J. Phys. 46 053001 (2025)}; doi: %
\href{https://doi.org/ 10.1088/1361-6404/adf9a5}
{ 10.1088/1361-6404/adf9a5}
\end{abstract}

\maketitle

\section{Introduction}
\label{sec:introduction}

The Dirac equation, conceived by Paul Dirac in 1928
\cite{DIRAC1928A,DIRAC1928B}, is one of the cornerstones of contemporary
physics, merging the principles of Special Relativity and Quantum Mechanics.
This equation not only predicts antiparticles but also contains the
Pauli equation in the non-relativistic limit, consequently incorporating
the concept of spin.
Revered as a milestone in scientific history \cite{Kragh1981}, it is
also distinguished by its beauty, as highlighted by Wilczek
\cite{FARMELO2002}:
``Of all the equations of physics, perhaps the most magical is the Dirac
equation''.
Moreover, the applications of Dirac theory span various domains,
including Quantum Field Theory \cite{ZEE2010,OHLSON2011}, Materials
Science \cite{Neto2009, Novoselov2005}, and Cosmology
\cite{Corchero2001, Tutusaus2017}.

Beyond its mathematical and physical significance, the equation garners
attention outside academia, prominently featured in scientific
popularization materials \cite{PBS}.
This widespread coverage underscores its emblematic importance within
modern physics. However, despite its stature, Dirac's model often
remains unexplored in undergraduate studies.
Hence, our proposal aims to bridge this gap by presenting a didactic
introduction to the equation, not claiming to make an original
contribution to Relativistic Quantum Mechanics or to the History of
Physics while doing so.

Similar efforts have been undertaken
\cite{Kragh1981,SMIRMOV2016,THIBES2022}, and our aim is to complement
them by emphasizing the historical context coupled with the comparison
between the wave equations.
Keeping in mind the didactic purpose of this work, we have included
information beyond the wave equations, such as a historical preamble and
insights and anecdotes from Dirac throughout the article.
While this may seem tangential, we believe these additional details,
besides arousing the reader's curiosity, may provide a deeper
understanding of the nuances of the man who formulated the equation.

This work is intended for undergraduate students who are fascinated by
the subject, but are not content with a purely conceptual approach.
Instead, we utilize mathematical tools commonly taught in the early
years of undergraduate education, ensuring that the article remains
accessible without posing significant challenges.
Although we aim to introduce the concepts progressively, a prior
familiarity with non-relativistic Quantum Mechanics is recommended,
particularly for subsections \ref{subsec:eletro_nr} and
\ref{subsec:fw_trans}.
For instances requiring more formal computation, we strive to clarify
the steps explicitly, accompanied by further recommended readings to aid
comprehension.
Consciously, we have chosen not to use the Dirac-Braket
notation, as it is typically not employed in the initial encounters with
Quantum Mechanics.
Exceptions are made in the appendix \ref{sec:Appendix}, where the
notation significantly facilitates the discussion regarding the
dimension of the Dirac matrices, and in subsection
\ref{subsec:fw_trans}, where detailed calculations would be quite 
extensive.

Furthermore, we seek to fill a gap in references that address the
incongruity between the Schr\"{o}dinger equation and Relativity
\cite{Kragh1981,RAJASEKARAN2003, PAIS1988}.
Surprisingly, even Relativistic Quantum Mechanics books
\cite{Bjorken1964,GREINER2000} may lack this discussion, usually
starting the detailed analysis from the Klein-Gordon (KG) equation
onward. 
To tackle this issue, we choose to demonstrate the equation's
non-invariance using a simple one-dimensional Lorentz transformation, a
key principle from Relativity Theory, explained as in introductory
Modern Physics classes \cite{TIPLER2012}.
We carry out this invariance test using the same methodology as the KG
equation so that the contrast between the models is favored.
This direct comparison between the wave equations is intensified so that
the reader can perceive the distinctive characteristics of each
equation.
Reference \cite{Levy1985} is an inspiration in this regard, as it draws
an analogous comparison between the models but is restricted to the more
complex case of atoms resembling hydrogen.

Moreover, we make use of covariant relativistic notation to highlight,
in a non-rigorous fashion, the invariance of the Dirac equation and
present more insightful versions of both the KG and Dirac equations.
We presume that the reader, interested in the Dirac equation, has
encountered its more refined form expressed through this
notation. Consequently, we contend that learning the relativistic
covariant notation is beneficial and fosters a deeper understanding of
the equation.

This paper is organized as follows.
Sec. \ref{sec:historical} provides a historical prelude to wave
mechanics, with emphasis on the development of matrix mechanics.
In Sec. \ref{sec:Schr\"{o}dinger}, we introduce the Schr\"odinger
equation and offer an overview of its key features.
Sec. \ref{sec:klein-gordon} presents the KG equation, focusing on its
relativistic foundations and comparing it to the Schr\"odinger equation.
Sec. \ref{sec:pauli} examines how Pauli introduced spin into the
description of electromagnetic interactions within the Schr\"odinger
equation framework.
The Dirac equation is discussed in Sec. \ref{sec:dirac}.
Subsections \ref{subsec:eletro_nr} and \ref{subsec:fw_trans} present two
approaches to derive its non-relativistic limit, while  section
\ref{sec:relativistic_notation} briefly introduces the covariant
relativistic notation.
Finally, Sec. \ref{sec:conc} summarizes the main points developed
throughout the paper.
The appendix \ref{sec:Appendix} discusses the dimension of the Dirac
matrices.

\section{Historical Preamble}
\label{sec:historical}

The development of the Dirac equation is intricately connected to the
progression of Quantum Mechanics and Dirac's professional
journey.
This section explores the historical context of quantum theory leading
to the introduction of wave mechanics, as we believe that understanding
the scope of quantum physics prior to introducing the initial wave
equation is vital to understanding the significance of the
Schr\"{o}dinger equation.
Recommended supplementary readings include \cite{Bernstein2005}, which
highlights Born's influence during this era and elaborates further on
matrix mechanics, and \cite{Gottfried2011}, which concentrates on
Dirac's role before the formulation of his equation.

On 21 September 1925, Werner Heisenberg published an exploratory article
titled \emph{Quantum-theoretical re-interpretation of kinematic and
  mechanical relations} \cite{HEISENBERG1925} in response to
discrepancies between experimental observations and theoretical
predictions regarding the spectral lines of the hydrogen emission
spectrum \cite{JAMMER1989}.
In this work, a significant transition was signaled, moving away from
the \emph{Old Quantum Mechanics} period.
This era included major breakthroughs in quantum physics between 1900
and 1925, including the de Broglie hypothesis, the Pauli exclusion
principle, and the Bohr-Sommerfeld atom model \cite{Pais1982}.
Despite successes, the Old Quantum Mechanics exhibited methodological
gaps, relying on an amalgamation of hypotheses, principles, and theorems
\cite{JAMMER1989}.
While Bohr endeavored to reconcile spectral line discrepancies through
his correspondence principle, Heisenberg diverged from such approaches
and classical methods, choosing instead to ground his theory on
measurable observables.
Fundamentally, he departed from non observable quantities within the
quantum scope.
For example, instead of using the position of an electron traversing an
orbit, one should utilize the probability amplitude $X_{nm}$ of the
particle undergoing a transition from state $n$ to state $m$
\cite{Bernstein2005}.
Based on this reasoning, he formulates his well-known
\emph{multiplication rule}, by which two observables, such as $A_{m k}$
and $B_{k n}$, are multiplied to form a new observable $(AB)_{m n}$,
which represents a combination of the original observables
\begin{equation}
        (AB)_{m n}=\sum_{k=0}^{\infty} A_{m k} B_{k n}.
        \label{multiplication_rule}
\end{equation}

A broad discussion of the mathematics of this topic is found in
\cite{JAMMER1989}. However, computing the product
\eqref{multiplication_rule} in reverse order, that is, $B_{m k}$ and
$A_{k n}$, generally produces a different combination, so that
$(AB)_{mn} \neq (BA)_{m n}$.
Remarkably, this differs from the classical theory, where the product of
two quantities (c-numbers), $a$ and $b$ is independent of the order,
i.e., $ab = ba$. 

In Heisenberg's algebraic framework, Born identified matrix properties,
a subject rarely utilized in physics until then
\cite{JAMMER1989}.
Consequently, one would infer that Heisenberg's observables,
representing physical quantities in Quantum Mechanics, took
the form of matrices. 
Born, along with Jordan, devised what would become known as matrix
mechanics in their 1925 publication, "On Quantum Mechanics"
\cite{BORN1925}, received merely six days after the publication of
Heisenberg's article in the same scientific journal
\emph{Zeitschrift für Physik}.
By applying the Heisenberg rule, they successfully derived the canonical
commutation relation in matrix form \cite{Bernstein2005}
\begin{equation}
  \sum_{k=0}^{\infty} \left(X_{n k} P_{k n}-P_{n k} X_{k n} \right) =
  \frac{h}{2\pi}i \mathbbm{1}_{n},
  \label{canonical_commutation_relation_matrices_2}
\end{equation}
where $h$ is Planck's constant, $\mathbbm{1}_{n}$ is the unity matrix in
dimensions $n \times n$ and $X_{n k}$ and $P_{k n}$ are matrices or
observables.
This asserts that it is impossible for states to measure both a definite
position and momentum simultaneously, a realization that is one of the
foundations of Quantum Mechanics.
In November 1926, a follow-up to the initial paper \cite{BORN1926A},
this time co-authored by Heisenberg, was published.
This publication further established the principles of matrix mechanics
with greater thoroughness and detail, and Born famously called it the
``three-man paper'' \cite{JAMMER1989}.

Dirac, in turn, noticed an initially perplexing relation in Heisenberg's
article that, weeks later, would lead to his first publication in
Quantum Mechanics: the non-commutation between certain pairs
of physical quantities \cite{PAIS1988}. 
This relationship underwent a period of contemplation before Dirac, on a
Sunday in October 1925, linked it to the Poisson brackets
\cite{KRAGH1990}, which are crucial components of Hamiltonian
formalism in Classical Mechanics.
In December 1925, \emph{The Fundamental Equations of Quantum Mechanics}
\cite{DIRAC1925} is published, wherein, without using Heisenberg
products or matrix properties, it was established the commutation
relation
\begin{equation}
  [x,p] =i\hbar,
  \label{canonical_commutation_relation}
\end{equation}
where $x$ and $p$ are what would soon be called operators, or, in Dirac
terms, "q-numbers", and $\hbar=h/2\pi=1.0546 \times 10^{-34}$ J.s is the
reduced Planck's constant.
Here
\begin{equation}
  [x,p]=xp-px,
  \label{canonical_commutation_relation_matrices}
\end{equation}
is the commutator between $x$ and $p$, which, when associated with
Poisson brackets, illustrates a connection between quantum theory and
the Hamiltonian formalism that Dirac aimed to achieve.
Dirac's aspiration originated from his conviction that classical
equations retained their validity, although they needed alterations in
their inherent mathematical properties \label{dirac_convictions}.
The process of quantizing classical theories, just as Dirac did with
Poisson brackets, is called \emph{canonical quantization}.

\begin{figure*}
    \centering
    \includegraphics[width=\textwidth]{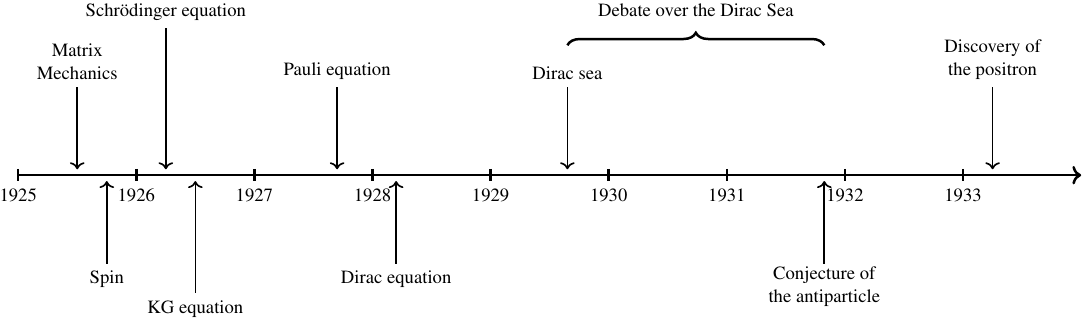}
    \caption{
      Not to scale timeline concerning the most relevant events for this
      work.
    }
    \label{fig:timeline}
\end{figure*}

Despite the independent deduction of this relationship, Dirac's work was
well received.
He was only 23 years old and still a student at Cambridge, so Born would
remark:
“The name Dirac was completely unknown to me.
The author seemed to be a young man, but everything was perfect
in his way and admirable” \cite{Born1978}.
Moreover, by offering a more coherent theoretical framework, matrix
theory was notably strong as Pauli employed matrix mechanics to derive
the Balmer formula for hydrogen atom spectra \cite{Pauli1926} -- albeit
with the aid of supplementary assumptions \cite{BELLER1983}.
However, the theory of Heisenberg, Born and Jordan was not immune to
criticism: there were problems with respect to the physical
interpretation of the theory, as well as a difficulty in defining a
stationary state \cite{BELLER1983}.
Furthermore, there was a lack of familiarity with the mathematics
involved and the concepts adopted. For instance, Fermi only attained a
comprehensive understanding of quantum theory through Schr\"{o}dinger's
wave mechanics \cite{JAMMER1989}.
The timeline in Fig. \ref{fig:timeline}, presented out of scale,
illustrates some of the most crucial events within the scope of this
article.

\section{Schr\"{o}dinger equation}
\label{sec:Schr\"{o}dinger}

In the first semester of 1926, while Dirac was pursuing his Ph.D., Erwin
Schr\"{o}dinger published four works on wave mechanics
\cite{SCHRöDINGER1926A,SCHRöDINGER1926B,SCHRöDINGER1926C,SCHRöDINGER1926D},
all titled \emph{Quantization as an eigenvalue problem}.
Schr\"{o}dinger's initial goal was to derive an equation for de
Broglie's matter waves and, to this end, he used Hamilton's
optical-mechanics analogy \cite{QUAGLIO2021, JAMMER1989}.
He proposed that classical mechanics represents a specific instance
within the broader scope of mechanics, encompassing microscopic
scenarios, a parallel to geometric optics within the wider domain of
optics.
Consequently, classical equations were as ineffective for quantum
problems as geometric optics was for wavelike problems.
Therefore, it was necessary to establish wave mechanics akin to physical
optics, with the wave equation replacing classical equations of motion.
In hindsight, this approach reflects Dirac's convictions about the
connections of Quantum Mechanics with the relations in classical
physics, as discussed in the preceding section.

The derivation of the Schr\"{o}dinger equation presented here emphasizes
some important aspects regarding the quantization framework, which will
be revisited throughout the reading.
We commence with the classical relationship between energy $E$ and
momentum, $\mathbf{p}$
\begin{equation}
  \label{classical_energy_momentum_relation}
  E=\frac{\mathbf{p^2}}{2m}+V(\mathbf{r},t).
\end{equation}
where $m$ is the mass of the classical particle and $V(\mathbf{r},t)$
the potential energy, depending on both the position vector $\mathbf{r}$
and  time $t$.
Schr\"{o}dinger proposed the following substitutions \cite{SchrödingerE}
\begin{align}
  E
  &
    \rightarrow i\hbar\frac{\partial}{\partial t},\label{energy_operator}\\
  \mathbf{p}
  &
    \rightarrow -i\hbar\boldsymbol{\nabla},\label{momentum_operator}
\end{align}
The relations above are a clear example of canonical quantization.
Their specific form stems from Hamilton's optical-mechanical analogy, in
which the motion of a particle in a conservative force field obeys the
same formal laws as the propagation of light through a refractive medium
\cite{JAMMER1989,QUAGLIO2021}.
In this context, they can be seen as expressing the connection between
Hamilton’s equations for a particle and the corresponding wave equation.
Additionally, the emergence of the quantum constant $\hbar$ is justified
by its dimensional equivalence to that of action
\cite{SCHRöDINGER1926A}.
The presence of the imaginary unit arises from a separation of variables
of the form $\psi(\mathbf{r},t) = u(\mathbf{r}) \exp(-i E t / \hbar)$
in the derivation of the time-dependent Sch\"odinger equation
\cite{SCHRöDINGER1926D}.

Replacements \eqref{energy_operator} and \eqref{momentum_operator}
symbolize an important step in Quantum Mechanics: the
transition of quantities to \emph{operators}, which are specific kinds
of linear transformation present in linear algebra.
Replacing the aforementioned operators in
\eqref{classical_energy_momentum_relation} we obtain
\begin{equation}
  \label{schrodinger_operator}
  i\hbar\frac{\partial}{\partial t}=-\frac{\hbar^2}{2m}\boldsymbol{\nabla}^2+V(\mathbf{r},t),
\end{equation}
where $\boldsymbol{\nabla}^{2}$ is the Laplacian differential operator
\begin{equation}
    \boldsymbol{\nabla}^{2}=\frac{\partial^2}{\partial x^2}+\frac{\partial^2}{\partial y^2}+\frac{\partial^2}{\partial z^2}.
\end{equation}

In Eq. \eqref{schrodinger_operator}, the operators appear on both
sides, acting on the states to acquire physical significance.
For now, the state undergoing the action of operators is the function
$\psi\left(\mathbf{r},t \right)$.
By applying both sides of \eqref{schrodinger_operator} to the state
$\psi \left(\mathbf{r},t \right)$, we obtain
\begin{equation}
  \label{schrodinger_equation}
  i\hbar\frac{\partial}{\partial t}\psi \left(\mathbf{r},t\right)
  =\left[-\frac{\hbar^{2}}{2m}\boldsymbol{\nabla}^{2}
    +V \left(\mathbf{r},t \right) \right]\psi \left(\mathbf{r},t \right),
\end{equation}
a linear partial differential equation, which is referred to as
time-dependent \emph{Schr\"{o}dinger equation}.
The function $\psi \left(\mathbf{r},t \right)$ is precisely the solution
of this differential equation, a complex valued function of the position
and time associated with the matter wave.

In Quantum Mechanics, the operator linked to energy is termed
Hamiltonian, denoted as $H$ from now onward, while $E$ will be used to
represent an energy quantity -- occasionally replaced by its
corresponding operator \eqref{energy_operator}.
Besides being proportional to the time derivative, the Hamiltonian can
be formulated in terms of other operators, depending on the specific
model under consideration.
For Eq. \eqref{schrodinger_equation}, it can be represented as
\begin{equation}
    H=-\frac{\hbar^{2}}{2m}\boldsymbol{\nabla}^{2}+V(\mathbf{r},t). \label{schrodinger_hamiltonian}
\end{equation}
Thus, the Hamiltonian is defined in such a way that it enables us to
express Eq. \eqref{schrodinger_equation} as
\begin{equation}
  i\hbar\frac{\partial}{\partial t}\psi \left(\mathbf{r},t \right)
  =H\psi \left(\mathbf{r},t \right),
  \label{generalized_schrodinger_equation}
\end{equation}
which is applicable, as we will see later, to other models beyond the
Schr\"{o}dinger one, and for this reason, it is called
\emph{generalized Schr\"{o}dinger equation}.

With the equation now introduced, we proceed to outline several
important aspects and merits of the Schr\"odinger equation.
We begin by discussing the motivation behind the titles of Schrödinger’s
works
\cite{SCHRöDINGER1926A,SCHRöDINGER1926B,SCHRöDINGER1926C,SCHRöDINGER1926D},
which present Quantum Mechanics as an eigenvalue problem.

The relationship between eigenvalues and quantum-mechanical operators is
distinctive, enabling the derivation of concrete physical outcomes.
When an operator is applied to a specific quantum state, defined by a
given eigenfunction, the result is the same state multiplied by a
scalar.
The eigenvalue approach of Schr\"{o}dinger consists precisely in
applying this aforementioned relation, using the Hamiltonian extracted
from his equation as an operator and making it act on the state
$\Psi \left(\mathbf{r},t \right)$, that is, an eigenfunction
of $H$.
The eigenvalues $\varepsilon$ arise as solutions to the equation
\begin{equation}
  H\Psi \left(\mathbf{r},t \right)=
  \varepsilon\Psi \left(\mathbf{r},t \right).
    \label{eigenvalue_psi}
\end{equation}
By employing this method, Schr\"{o}dinger notably succeeded in deriving
the energy levels that correspond to Bohr's non-relativistic model for
the Hydrogen atom \cite{SCHRöDINGER1926A,JAMMER1989}.

Having established the utility of the wave function, we now turn our
attention to the nature of $\psi\left(\mathbf{r},t \right)$.
Sch\"{o}dinger \cite{SCHRöDINGER1926D} had an
electrodynamic interpretation of his wave equation, assuming that the
square modulus of  $\psi \left(\mathbf{r},t \right)$ was a function of
the charge distribution.
However, in June 1926, Born introduced his probabilistic interpretation
\cite{BORN1926B,PAIS1988}, considering the square modulus of
$\psi \left(\mathbf{r},t \right)$ as a probability density
\begin{equation}
    \rho(\mathbf{r},t)=\left|\psi \left(\mathbf{r},t \right)\right|^{2},\label{rho_density}
\end{equation}
where $\rho(\mathbf{r},t)$ is always positive. Later in this section,
when we discuss the free particle, we will exemplify this
interpretation.

\begin{figure}[b]
  \centering
  \includegraphics[width=\columnwidth]{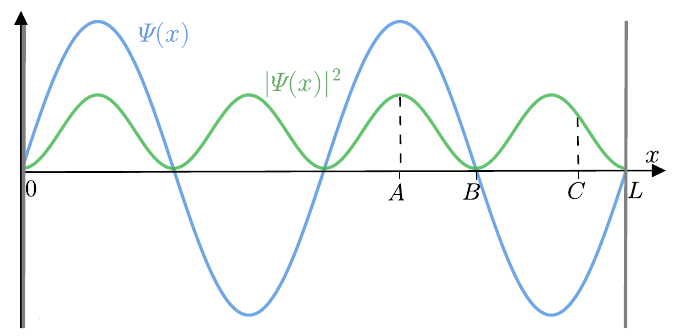}
  \caption{
    Schematic representation of $\psi(x)$ and $|\psi(x)|^2$ for the
    scenario of a particle in a box of length L, a simple solution to
    the Schr\"{o}dinger equation.
  }
  \label{fig:probabilistic_interpretation}
\end{figure}

By manipulating \eqref{schrodinger_equation}, its conjugate, and
\eqref{rho_density}, we can derive the \emph{equation of continuity}
\begin{equation}
  \frac{\partial}{\partial t}\rho(\mathbf{r},t)
  + \boldsymbol{\nabla}\cdot{\mathbf{j}(\mathbf{r},t)}
  =0,
  \label{continuity_equation}
\end{equation}
where $\mathbf{j(\mathbf{r},t)}$ is the \emph{probability current},
defined as
\begin{equation}
  \mathbf{j}(\mathbf{r},t)=
  -\frac{i\hbar}{2m}
  \left[
    \psi^{*}\boldsymbol{\nabla}\psi-(\boldsymbol{\nabla}\psi^{*})\psi
  \right],
  \label{j_current}
\end{equation}
where we leave the position and time variables of
$\psi \left(\mathbf{r},t \right)$ implicit.
Within the frameworks of Electromagnetism and Fluid Dynamics,
Eq. \eqref{continuity_equation} represents the principle of charge
conservation and the conservation of fluid mass, respectively.
Concerning the Schr\"{o}dinger equation, it signifies
\emph{conservation of probability}.

It is natural to wonder how wave mechanics fits in with matrix
mechanics.
In his third article \cite{SCHRöDINGER1926C}, Schr\"{o}dinger
established that wave mechanics implies some concepts from matrix
mechanics.
John von Neumann, only in 1929, formally demonstrated the
equivalence between matrix mechanics and wave mechanics, based on a
functional analysis theorem \cite{JAMMER1989}.

Next, we focus on a essential application of the Schr\"odinger equation:
the case of a free particle.
In this scenario, we set $V \left(\mathbf{r},t \right)=0$ in Eq.
\eqref{schrodinger_equation} and rearrange the terms, solving for an
eigenfunction $\Psi \left(\mathbf{r},t \right)$, such that
\begin{equation}
  \label{schrodinger_equation_2}
  \frac{\hbar^{2}}{2m}\boldsymbol{\nabla}^{2}
  \Psi \left(\mathbf{r},t \right)
  + i\hbar\frac{\partial}{\partial t}\Psi \left(\mathbf{r},t\right)=0,
\end{equation}
which is akin to the plane wave solution found in Maxwell's
equations within a vacuum
\begin{equation}
  \boldsymbol{\nabla}^{2}\mathbf{E}(\mathbf{r},t)
  -\frac{1}{c^{2}}\frac{\partial^{2}}{\partial^{2}t}\mathbf{E}(\mathbf{r},t)=0.
  \label{maxwell_wave_equation}
\end{equation}
A notable distinction is the presence of the second-order time
derivative in the latter, which is absent in the former. This aspect
will be addressed in a subsequent discussion.

Still within the context of the free particle, we consider the
particle-in-a-box case, a concrete example of energy quantization
resulting from boundary conditions.
We consider the one-dimensional and time-independent case, with the
associated particle confined in a box of length $L$, the Schr\"{o}dinger
equation admits sinusoidal solutions $\Psi(x)$, as
illustrated in Fig. \ref{fig:probabilistic_interpretation}, with
positions $A$, $B$, and $C$ taken as examples.
According to the magnitude of the squared modulus
\eqref{rho_density_dirac_equation}, it can be observed that the
decreasing order of probability for these positions is $A$, $C$, and
$B$.
The reader more familiar with Quantum Mechanics recognizes
that, due to the probabilistic interpretation, the integration of
$\rho(\mathbf{r},t)$ over the length of the box (from $x=0$ to $x=L$)
yields unity.
This procedure, commonly found in textbook problems, usually seeks to
identify the normalization constant of the specified wave function.
The energy eigenvalues for this case are obtained through the equation
\begin{equation}
    \varepsilon_n=\frac{\hbar^2 \pi^2}{2 m L^2} n^2, \quad n=1,2,...,
\end{equation}
from where we extract what seems to be a prerequisite, but later we will
discover that it is not: positive definite energy eigenvalues. 
We further note that a given state $\psi(x)$ can be expressed as a
linear combination of the eigenfunctions $\Psi(x)$, as a consequence of
the linearity of the Schrödinger equation.

The scientific community received the Schr\"{o}dinger equation with
enthusiasm, and prominent personalities such as Fritz London, Charles
Darwin, and Enrico Fermi were among those who welcomed it.
Its emergence was almost reassuring in the context of the departure from
classical methods advocated by the newer generation of physicists
\cite{FARMELO2002}.
At the Munich Conference in the summer of 1926, most participants
aligned themselves with wave mechanics to the detriment of matrix
mechanics \cite{BELLER1983}.

On the other hand, Dirac had an initial aversion to Schr\"{o}dinger's
equation \cite{PAIS1988}, since he had already developed his own
methodology to address quantum problems involving the aforementioned
q-numbers.
However, he gradually accepted the Schr\"{o}dinger equation as a
complement to his theory and applied wave mechanics to problems with
many particles \cite{DIRAC1926}; in the same paper, he introduced his
\emph{transformation theory}.
One characteristic of this theory is that it generalizes Born's
probabilistic interpretation of the Schr\"{o}dinger equation
\cite{DIRAC1927}.

The merits of the Schr\"{o}dinger wave equation are vast and are widely
covered in the literature
\cite{SAKURAI2020,GRIFFITHS2018,FEYNMAN2015,EISBERG1985}.
Planck would say that it
``plays the same role in modern physics as the equations established by
Newton, Lagrange, and Hamilton in classical mechanics'' 
\cite{PLANCK1931}.
However, for the purposes of this article, we will address its
limitations.

Feynman asserted that the Schr\"{o}dinger equation is "capable of
explaining all atomic phenomena, except those involving magnetism and
relativity" \cite{FEYNMAN2015}.
This prompts the question: why does the equation not conform to
Relativity? A physical model is considered compatible with Relativity
when it remains invariant under Lorentz transformation, a linear
transformation that converts coordinates from one reference frame to
another, moving at a constant speed relative to the first, replacing the
Galilean transformation of Newtonian physics.
This transformation serves as a tool to assess whether a given equation
aligns with the principles of Relativity.
A notable example of invariance is found in the solution of Maxwell's
equations for a vacuum \eqref{maxwell_wave_equation}, where the orders
of spatial (Laplacian) and temporal derivatives are equal.
On the other hand, equation \eqref{schrodinger_equation}, due to the
different orders of derivatives, intuitively suggests a lack of
``equality'' between space and time as proposed by Einstein's theory;
this is the brief explanation provided by the majority of textbooks
\cite{Kragh1981,RAJASEKARAN2003,PAIS1988}.

\begin{figure}
    \centering
    \includegraphics[width=0.85\columnwidth]{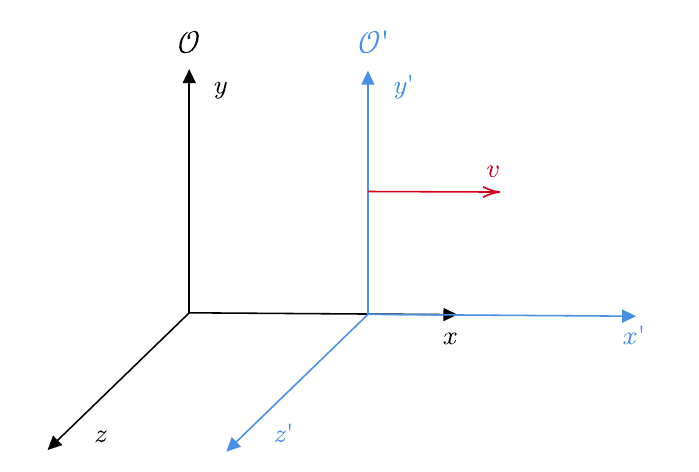}
    \caption{
      The inertial frames $\mathcal{O}$ and $\mathcal{O}'$.
      }
    \label{fig:Inertial frames}
\end{figure}

We will demonstrate that intuition is correct through a somewhat
informal one-dimensional deduction.
Therefore, we consider that in a reference frame $\mathcal{O}$, our wave
function is $\psi(x,t)$, and in a reference frame $\mathcal{O^{\prime}}$
moving with constant speed $v$ relative to $\mathcal{O}$, the wave
function is $\psi'\left(x',t' \right)$.
Figure \ref{fig:Inertial frames} illustrates the reference frames, and
the Lorentz transformation, in this case, tells us how the position ($x$
and $x'$) and the time ($t$ and $t'$) in the two reference frames are
related.

The structure of the equation must remain unchanged during a coordinate
transformation to maintain the covariance of the physical laws governed
by it.
The solutions in different reference frames assume distinct functional
forms which justify our writing of $\psi'(x',t') =
\psi(x(x',t'),t(x',t'))$.
We start with the Schr\"{o}dinger equation for the wave function in the
reference frame $O$ in the form
\begin{equation}
  i\hbar\frac{\partial}{\partial t}\psi \left(x,t \right)=
  -\frac{\hbar^{2}}{2m}
  \frac{\partial^{2}}{\partial x{{}^2}}\psi\left(x,t \right)
  +V(x,t)\psi \left(x,t \right),
  \label{desired_schrodinger_equation_another_frame}
\end{equation}
and perform the transformation from $\mathcal{O}$ to $\mathcal{O}'$
searching for the equation that $\psi'(x',t') = \psi(x(x',t'),t(x',t'))$
satisfies for the potential $V'(x',t') = V(x(x',t'),t(x',t'))$.
The one-dimensional Lorentz transformation in this case is
\begin{align}
  x' = {} & \gamma \left(x-vt \right),\label{x_another_frame}\\
  t' = {} & \gamma \left(t-\frac{v}{c^{2}}x \right),
             \label{t_another_frame}
\end{align}
where $\gamma$ is the Lorentz factor
\begin{equation}
  \gamma=\left(\sqrt{1-\frac{v^{2}}{c^{2}}}\right)^{-1}.
  \label{lorentz_factor_gamma}
  \end{equation}

Analyzing the Schr\"{o}dinger equation \eqref{schrodinger_equation}, it
is clear that to find the equation satisfied by
$\psi' \left(x',t'\right)$, we must first write $\partial/\partial t$
and $\partial^{2}/\partial x^{2}$ in terms of $\partial/\partial t'$ and
$\partial/\partial x'$.
Applying the chain rule, we have
\begin{align}
  \frac{\partial}{\partial x} = {}
  &
    \frac{\partial x'}{\partial x}\frac{\partial}{\partial x'}
    +\frac{\partial t'}{\partial x}\frac{\partial}{\partial t'},\nonumber \\
  = {}
  &
    \gamma\frac{\partial}{\partial x'}
    -\gamma\frac{v}{c{{}^2}}\frac{\partial}{\partial t'}. \label{partial_x_another_frame}
\end{align}
The second derivative as a function of $x$ is, again with the chain rule
\begin{align}
  \frac{\partial^{2}}{\partial x^{2}} = {}
  & \frac{\partial x'}{\partial x}\frac{\partial}{\partial x'}\left(\gamma\frac{\partial}{\partial x'}-\gamma\frac{v}{c{{}^2}}\frac{\partial}{\partial t'}\right)+\frac{\partial t'}{\partial x}\frac{\partial}{\partial t'}\left(\gamma\frac{\partial}{\partial x'}-\gamma\frac{v}{c{{}^2}}\frac{\partial}{\partial t'}\right),\nonumber \\
  ={}
  & \gamma\frac{\partial}{\partial x'}\left(\gamma\frac{\partial}{\partial x'}-\gamma\frac{v}{c{{}^2}}\frac{\partial}{\partial t'}\right)-\gamma\frac{v}{c{{}^2}}\frac{\partial}{\partial t'}\left(\gamma\frac{\partial}{\partial x'}-\gamma\frac{v}{c{{}^2}}\frac{\partial}{\partial t'}\right),\nonumber \\
  = {}
  & \gamma^{2}\left(\frac{\partial^{2}}{\partial x'^{2}}-2\frac{v}{c{{}^2}}\frac{\partial^{2}}{\partial x'\partial t'}+\frac{v^{2}}{c^{4}}\frac{\partial^{2}}{\partial t'^{2}}\right). \label{second_order_partial_x_another_frame}
\end{align}
Repeating the same procedure as above for the first order derivative of
$t$, we obtain
\begin{align}
  \frac{\partial}{\partial t}= {}
  &
    \frac{\partial x'}{\partial t}\frac{\partial}{\partial x'}
    +\frac{\partial t'}{\partial t}\frac{\partial}{\partial t'},\nonumber \\
  = {}
  &
    -\gamma v\frac{\partial}{\partial x'}
    +\gamma\frac{\partial}{\partial t'}.
    \label{partial_t_another_frame}
\end{align}

We will not consider the second derivative of time in Schr\"{o}dinger
equation for now, but we will deduce the expression as it will be useful
throughout the next section
\begin{align}
  \frac{\partial{{}^2}}{\partial t{{}^2}}  = {}
  &
  \frac{\partial x'}{\partial t}\frac{\partial}{\partial x'}\left(-\gamma v\frac{\partial}{\partial x'}+\gamma\frac{\partial}{\partial t'}\right)+\frac{\partial t'}{\partial t}\frac{\partial}{\partial t'}\left(-\gamma v\frac{\partial}{\partial x'}+\gamma\frac{\partial}{\partial t'}\right),\nonumber \\
  ={}
  & -\gamma v\frac{\partial}{\partial x'}\left(-\gamma v\frac{\partial}{\partial x'}+\gamma\frac{\partial}{\partial t'}\right)+\gamma\frac{\partial}{\partial t'}\left(-\gamma v\frac{\partial}{\partial x'}+\gamma\frac{\partial}{\partial t'}\right),\nonumber \\
   = {}
  &
    \gamma{{}^2}v{{}^2}\frac{\partial{{}^2}}{\partial x'{{}^2}}-\gamma{{}^2}v\frac{\partial}{\partial x'\partial t'}-\gamma{{}^2}v\frac{\partial}{\partial t'\partial x'}+\gamma{{}^2}\frac{\partial{{}^2}}{\partial t'{{}^2}},\nonumber \\
  = {}
  &
    \gamma{{}^2}\left(v{{}^2}\frac{\partial{{}^2}}{\partial
    x'{{}^2}}-2v\frac{\partial}{\partial x'\partial
    t'}+\frac{\partial{{}^2}}{\partial t'{{}^2}}\right).
    \label{second_order_partial_t_another_frame}
\end{align}
One can use Eqs. \eqref{second_order_partial_x_another_frame} and
\eqref{second_order_partial_t_another_frame} into the electromagnetic
wave equation, Eq. \eqref{maxwell_wave_equation}, considering the
one-dimensional case, and see that it is truly invariant.
On the other hand, doing the same in Eq.  \eqref{schrodinger_equation}
we have
\begin{align}
\!\!\!\!i\hbar & \left(-\gamma v\frac{\partial}{\partial x'}+\gamma\frac{\partial}{\partial t'}\right)\psi' \left(x',t' \right)-V'\left(x',t' \right)\psi' \left(x',t' \right) \notag\\
\!\!\!\!& =-\frac{\hbar^{2}}{2m}\left[\gamma^{2}\left(\frac{\partial^{2}}{\partial x'^{2}}-2\frac{v}{c{{}^2}}\frac{\partial^{2}}{\partial x'\partial t'}+\frac{v^{2}}{c^{4}}\frac{\partial^{2}}{\partial t'^{2}}\right)\right]\psi' \left(x',t' \right).\label{schrodinger_equation_another_frame}
\end{align}
Thus, comparing with
Eq. \eqref{desired_schrodinger_equation_another_frame}, it becomes
evident that the Schr\"{o}dinger equation is not invariant under Lorentz
transformation.
Therefore, it is not proper when applied to relativistic particles.

Someone might wonder, out of interest, how the equation remains valid in
a non-relativistic parallel scenario: Is it invariant under Galilean
transformation?
In a one-dimensional Galilean transformation, the relation between
position and time in the two reference frames is given by
\begin{align}
    x' & =x-vt,\\
    t' & =t,
\end{align}
so that the first-order spatial derivative is
\begin{align}
  \frac{\partial}{\partial x} = {}
  &
    \frac{\partial x'}{\partial x}\frac{\partial}{\partial x'}
    +\frac{\partial t'}{\partial x}\frac{\partial}{\partial t'},\nonumber \\
  = {}
  &
    \frac{\partial}{\partial x'},
\end{align}
and the second order is
\begin{equation}
    \frac{\partial^{2}}{\partial x^{2}}=\frac{\partial^{2}}{\partial x'^{2}}. \label{second_order_x_another_frame_galilean}
\end{equation}
Furthermore, the time derivative is
\begin{align}
  \frac{\partial}{\partial t} = {}
  & \frac{\partial x'}{\partial
   t}\frac{\partial}{\partial x'}
    +\frac{\partial t'}{\partial t}\frac{\partial}{\partial t'},\nonumber \\
    ={} & -v\frac{\partial}{\partial x'}+\frac{\partial}{\partial t'}. \label{partial_t_another_frame_galilean}
\end{align}
Inserting Eqs. \eqref{second_order_x_another_frame_galilean} and
\eqref{partial_t_another_frame_galilean} into Eq.
\eqref{schrodinger_equation}, we obtain
\begin{align}
  i\hbar \left(-v\frac{\partial}{\partial x'}
  +\frac{\partial}{\partial t'}\right)\psi'
  \left(x',t' \right) {} =
  &
    -\frac{\hbar^{2}}{2m}\frac{\partial^{2}}{\partial x'^{2}}\psi'
    \left(x',t' \right)\notag \\
  & +V'\left(x',t' \right)\psi' \left(x',t' \right),
\end{align}
which can be written as
\begin{align}
  i\hbar\frac{\partial}{\partial t'}\psi' \left(x',t' \right) = {}
  &
    -\frac{\hbar^{2}}{2m}\left(\frac{\partial^{2}}{\partial x'^{2}}-i\hbar v\frac{\partial}{\partial x'}\right)\psi' \left(x',t' \right)\notag \\ &+V'\left(x',t' \right)\psi \left(x',t' \right).\label{sr}
\end{align}
The second term on the right-hand side of Eq. (\ref{sr}) indicates that
the Schr\"{o}dinger equation is not invariant under the Galilean
transformation.
It is necessary to add a phase so that the wave function remains
consistent \cite{Padmanabhan2011}, and the probabilities do not change
\cite{HOME1997}.

Initially, Schr\"{o}dinger himself attempted to derive a valid
relativistic equation, but was unsuccessful in calculating the energy
eigenvalues \cite{JAMMER1989} according to the experimentally
successful Sommerfeld energy levels, derived from the quantization of
the \emph{relativistic} Bohr atom.
The Sommerfeld expression \cite{SOMMERFELD1916} for the energy
eigenvalues of the hydrogen atom is
\begin{equation}
  \varepsilon_{n,k}=
  \frac{mc^{2}}{\sqrt{1+\frac{\alpha^{2}}{\left(n-k-\sqrt{k^{2}-\alpha^{2 }}\right)^{2}}}}-mc^{2},\label{sommerfeld_formula}
\end{equation}
where $c$ means the speed of light in vacuum, $\alpha$ is the fine
structure constant, $n$ corresponds to the principal quantum number and
$k$ represents the azimuthal quantum number.
Sommerfeld's relativistic correction is associated with the splitting of
hydrogen spectral lines, known as the \emph{fine structure} of hydrogen.

In Eq. \eqref{sommerfeld_formula}, Schr\"{o}dinger obtained $n-k+1/2$
and $k-1/2$ instead of $n-k$ and $k$, respectively
\cite{Kragh1981}. However, it is known that this error is due to the non
incorporation of electron spin into the equation \cite{JAMMER1989}.
Schr\"{o}dinger eventually informed Dirac that he had developed a
relativistic wave equation, but it was unable to reproduce the
Sommerfeld formula.
In particular, Dirac displayed a noteworthy perspective, stating that
the colleague should have maintained confidence in his ``beautiful
relativistic theory'', even if it was inconsistent with accurate
experimental data \cite{KRAGH1990}.
A more in-depth discussion of this aspect of Schr\"{o}dinger's theory
can be found in \cite{Barley2021}.

\section{Klein-Gordon equation}
\label{sec:klein-gordon}

During 1926, a possible relativistic wave equation was derived by at
least seven authors: Klein \cite{KLEIN1926}, Schr\"{o}dinger
\cite{SCHRöDINGER1926D}, Fock \cite{FOCK1926}, Donder, Dungen
\cite{DONDER1926}, Kudar \cite{KUDAR1926} and Gordon \cite{GORDON1926}.
Priority in the nomenclature is given to Oscar Klein, who proposed the
equation in April 1926, while Fock provided a more intricate exploration
of relativistic wave mechanics \cite{Kragh1981}.
We will follow the more direct derivation, starting from the
relativistic relation between energy $E$ and linear momentum
$\mathbf{p}=(p_1,p_2,p_3)$, expressed as
\begin{equation}
    E^{2}=(\mathbf{p}c)^{2}+(mc^{2})^{2},
    \label{relativistic_energy}
\end{equation}
where the potential energy $V(\mathbf{r},t)$ is taken as zero and $m$
refers to the rest mass.
We can replace both $E$ and $\mathbf{p}$ quantities in
Eq. \eqref{relativistic_energy} by operators the operator in
Eqs. \eqref{energy_operator} and \eqref{momentum_operator}, so that we
obtain
\begin{equation}
    \frac{1}{c^{2}}\frac{\partial^{2}}{\partial t^{2}}=\boldsymbol{\nabla}^{2} -\left(\frac{mc}{\hbar}\right)^{2},
    \label{kg_operator}
\end{equation}
that we can apply the both sides to $\psi \left(\mathbf{r},t \right)$ to
arrive at the equation
\begin{equation}
    \frac{1}{c^{2}}\frac{\partial^{2}}{\partial t^{2}}\psi \left(\mathbf{r},t \right)=\boldsymbol{\nabla}^{2}\psi \left(\mathbf{r},t \right)-\left(\frac{mc}{\hbar}\right)^{2}\psi \left(\mathbf{r},t \right),
    \label{kg_expanded_equation}
\end{equation}
which can be written as
\begin{equation}
    \left[\frac{1}{c^{2}}\frac{\partial^{2}}{\partial t^{2}}-\boldsymbol{\nabla}^{2}+\left(\frac{mc}{\hbar}\right)^{2} \right]\psi \left(\mathbf{r},t \right)=0.
    \label{kg_equation_lhs}
\end{equation}
Moreover, employing the d'Alembertian operator
$\square=1/c^2\partial^2/\partial t^2 -\nabla^{2}$
and the constant $\mu={mc}/{\hbar}$, equation (\ref{kg_equation_lhs})
can be put in the form
\begin{equation}
    \left(\square+\mu^2 \right)\psi \left(\mathbf{r},t \right)=0,
    \label{kg_equation_with_dalembertian}
\end{equation}
which is called \emph{scalar wave equation} or \emph{Klein-Gordon
  equation} (KG from now on).
There is another way to write it, using relativistic covariant notation
and natural units, which we will show later.

The KG equation aims to adhere to the principles of Relativity,
prompting an initial assessment of its conformity from this perspective.
Notably, one aspect that immediately draws attention is the similarity
between equation \eqref{kg_equation_lhs} and the electric field wave
equation \eqref{maxwell_wave_equation} -- in fact, one can obtain the
former from the latter, as shown by \cite{TRETYAKOV2010,ELGAYLANI2014}.
Furthermore, we analyze the orders of the space and temporal
derivatives, and we see that they are equal: this leads to the intuition
that the KG equation is invariant under a Lorentz transformation.
A parallel analysis, akin to our examination of the Schr\"{o}dinger
equation, is now applied to the one-dimensional form of the KG equation,
i.e.,
\begin{equation}
    \left[\frac{1}{c^{2}}\frac{\partial^{2}}{\partial t^{2}}-\frac{\partial^{2}}{\partial x{{}^2}}+\left(\frac{mc}{\hbar}\right)^{2}\right]\psi(x,t)=0,
\end{equation}
in which we incorporate
Eqs. \eqref{second_order_partial_x_another_frame} and
\eqref{partial_t_another_frame}, we arrive at
\begin{align}
    &\left[\frac{\gamma^{2}v^{2}}{c^{2}}\frac{\partial^{2}}{\partial {x'}^{2}}+\frac{\gamma^{2}}{c^{2}}\frac{\partial^{2}}{\partial {t'}^{2}}-\gamma^{2}\frac{\partial^{2}}{\partial x'^{2}}-\gamma^{2}\frac{v^{2}}{c^{4}}\frac{\partial^{2}}{\partial t'^{2}}\right]\psi' \left(x',t' \right) \notag\\
&+\left(\frac{mc}{\hbar}\right)^{2}\psi' \left(x',t' \right) =0,
\end{align}
so that we rewrite it as
\begin{equation}
    \left[-\frac{\partial{{}^2}}{\partial x'{{}^2}}\gamma^{2}\left(1-\frac{v^{2}}{c^{2}}\right)+\frac{\partial{{}^2}}{\partial t'{{}^2}}\frac{\gamma^{2}}{c^{2}}\left(1-\frac{v^{2}}{c^{2}}\right)+\left(\frac{mc}{\hbar}\right)^{2}\right]\psi' \left(x',t' \right) =0,
\end{equation}
where we can employ the definition of the Lorentz factor \eqref{lorentz_factor_gamma}, so that
\begin{equation}
    \left[\frac{1}{c^{2}}\frac{\partial{{}^2}}{\partial t'{{}^2}}-\frac{\partial{{}^2}} {\partial x'{{}^2}}+\left(\frac{mc}{\hbar}\right)^{2}\right]\psi' \left(x',t' \right)=0,
\end{equation}
demonstrating that the KG equation remains invariant under Lorentz
transformation.

Another crucial aspect of the KG equation is investigating its
applicability in physical systems, particularly considering the range of
problems that can be addressed with the Schr\"{o}dinger equation.
Therefore, applying the KG equation to a simple system and testing its
validity is only fair.
For instance, as an elementary problem, let us analyze the energy
spectrum of a free particle.
Taking Eq. \eqref{kg_equation_lhs} and simplifying it, we have
\begin{equation}
    \frac{1}{c^2}\frac{\partial^{2}}{\partial t^{2}}\Psi \left(\mathbf{r},t \right)-\boldsymbol{\nabla}^{2}\Psi \left(\mathbf{r},t \right)+\mu {^2}\Psi \left(\mathbf{r},t \right)=0,\label{kg_equation_rewritten}
\end{equation}
which is expected to have plane wave solutions with a time
dependency of the type $-(i\varepsilon t)/\hbar$ and a spatial
dependence of the type $+(i\mathbf{p}\cdot\mathbf{r})/\hbar$, so
that
\begin{equation}
    \Psi \left(\mathbf{r},t \right)=N\exp[-\frac{i}{\hbar}(\varepsilon t-\mathbf{p}\cdot\mathbf{r})],
    \label{psi_wave_equation}
\end{equation}
where $N$ is a normalization constant and $\cdot$ symbolizes the scalar
product.
Equation \eqref{psi_wave_equation} is a solution of Eq.
\eqref{kg_equation_rewritten} if
\begin{equation}
    \varepsilon^{2}=\mathbf{p}^{2}c^2+\mu^{2}c^{2}\hbar^2,
\end{equation}
which includes both positive and negative energy eigenvalues.
This may seem counterintuitive, but it is now understood that negative
energy solutions are valid.
However, when the concept of antiparticles was unknown, this posed an
obstacle to accepting the KG equation.
In the next section, we will discuss this intriguing aspect, which
arises as a solution to the free particle.

Additionally, an important aspect to study is how the probability
density $\rho(\mathbf{r},t)$.
It is desirable that $\rho(\mathbf{r},t)$ be conserved and positive
definite in this model, as it happens in the Schr\"{o}dinger equation.
In this manner, we obtain the following equation
for the probability density
\begin{equation}
    \rho(\mathbf{r},t) =\frac{i\hbar}{2mc^2}\left(\psi^{*}\frac{\partial\psi}{\partial t}-\psi\frac{\partial\psi^{*}}{\partial t}\right),
\end{equation}
where we note that, although it is a preserved quantity,
$\rho(\mathbf{r},t)$ is not positive definite, unlike
Eq. \eqref{rho_density}.
This posed an immediate challenge as it complicated the probabilistic
interpretation of the wave function.
Nevertheless, we now know that the correct interpretation is that
$\rho(\mathbf{r},t)$ is a probability charge density
\cite{SAKURAI2020}.

Initially, the KG equation was considered the correct relativistic
generalization of wave mechanics:
It possessed the inherent elegance of symmetry and, thus, was invariant
under the Lorentz transformation.
Schr\"{o}dinger himself would use it in subsequent approaches to the
Compton effect, a phenomenon that encompasses relativistic effects.
However, despite its mathematical appeal, its applications were limited
compared to the Schr\"{o}dinger equation, especially considering the
lack of widely accepted explanations for negative energies.
This meant that, even though some physicists believed that the problem
of relativistic generalization was already solved with the KG equation,
the quest for a more suitable wave equation persisted.
However, it should be noted that in quantum field theory, the KG
equation finds validity for pions, which are spinless particles
\cite{OHLSON2011}.
Additionally, it is instructive to note that in the non-relativistic
limit, the KG equation reduces to the Schr\"{o}dinger equation
\cite{FESBACH1958}.

\section{Pauli equation}
\label{sec:pauli}

\begin{figure*}
  \centering
  \includegraphics[width=0.55\textwidth]{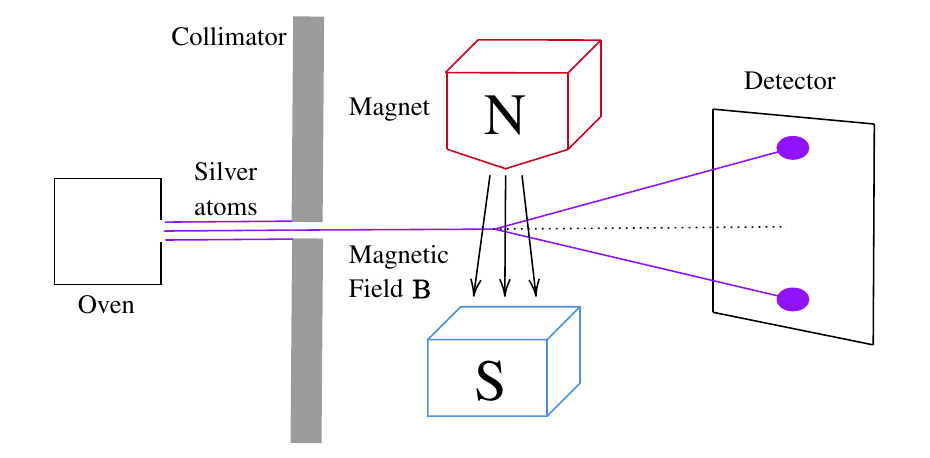}
  \caption{
    SG experiment illustrated schematically.
    The beam of silver atoms is expelled from the furnace, passes
    through a collimator, and is subjected to a heterogeneous magnetic
    field $\mathbf{B}$.
    In the detector, two preferential regions are perceived.
    }
  \label{fig:sg}
\end{figure*}

An important event in the search for a relativistic wave equation was
the discovery of spin, initially believed to be intrinsically
relativistic.
The first indication of spin's existence arose from the Stern-Gerlach
(SG) experiment \cite{STERN1922}, where an atomic beam of silver was
subjected to an inhomogeneous magnetic field. Silver's 47th electron's
magnetic moment approximates the atom's magnetic moment
because of its specific structure.
If the electron behaved classically, its magnetic moment would exhibit a
continuous range of values, resulting in a Gaussian pattern after
passing through the magnetic field.
However, in the SG experiment, deviations were observed for all atoms,
with detection occurring in two preferential regions, as illustrated in
Fig. \ref{fig:sg}. This suggested that the magnetic moment of the
electron had only two directions, a phenomenon which Sommerfeld
interpreted as evidence of space quantization \cite{JAMMER1989}.

Pauli began the study of this paradigm by examining the hyperfine
structure of the hydrogen atom, proposing that the nucleus possesses an
angular momentum that does not simply ``disappear'' \cite{PAULI1924}.
In November 1925, Dutch physicists Uhlenbeck and Goudsmit expanded this
notion by offering a more thorough explanation of this novel quantum
property \cite{UHLENBECK1925}.
They introduced a new quantum number associated with an extra degree of
freedom connected to the electron's magnetic moment, called spin.
This quantum number had two observable states in the SG experiment:
spin up and spin down.
The recognition of spin highlighted its essential
role in quantum theory.

First, let us discuss how the Schr\"{o}dinger equation deals with
electromagnetic interactions by writing it as a function of
$\mathbf{p}$ as
\begin{equation}
    i\hbar\frac{\partial}{\partial t}\psi \left(\mathbf{r},t
    \right)=\left[\frac{\mathbf{p^2}}{2m}
      +V(\mathbf{r},t)\right]\psi \left(\mathbf{r},t \right),\label{schrodinger_equation_p}
\end{equation}
We address electromagnetic interactions involving an electron by
introducing the canonical momentum
\begin{equation}
    \boldsymbol{\Pi}=\mathbf{p}-\frac{e}{c} \mathbf{A},
    \label{canonical_momentum}
\end{equation}
and performing the following substitutions, which is called
\emph{minimal coupling},
\begin{align}
    \mathbf{p} & \rightarrow \boldsymbol{\Pi}=\mathbf{p}-\frac{e}{c}\mathbf{A}, \\
    V(\mathbf{r},t) & \rightarrow V(\mathbf{r},t)+e\phi, \label{minimal_coupling_substituition}
\end{align}
where $e$ represents the elementary charge, $\mathbf{A}$ is magnetic
vector potential  and $\phi$  is the scalar electric potential.
Performing the substitutions and in the absence of a potential
$V(\mathbf{r},t)$ we rewrite Eq.  \eqref{schrodinger_equation_p} as
\begin{equation}
  i\hbar\frac{\partial}{\partial t}\psi \left(\mathbf{r},t \right)
  =\left[
    \frac{1}{2m}\left(\mathbf{p}-\frac{e}{c}\mathbf{A}\right)^{2}
    +e\phi \right]\psi \left(\mathbf{r},t \right).\label{schrodinger_equation_minimal_coupling}
\end{equation}
This final result does not take the spin into account.

In 1927, Wolfgang Pauli \cite{PAULI1927} approached this problem: he
introduced \emph{ad hoc}, a term in
\eqref{schrodinger_equation_minimal_coupling}.
This term accounted for the interaction between the particle's spin and
an external electromagnetic field, leading to the derivation of what
came to be known as the Pauli equation, which is given by
\begin{equation}
  i\hbar\frac{\partial}{\partial t}\psi \left(\mathbf{r},t \right)
  =\left\{ \frac{1}{2m}\left[\mathbf{\boldsymbol{\sigma}}\
      \cdot \left(\mathbf{p}
        -\frac{e}{c}\mathbf{A}\right)\right]^{2}
    +e\phi\right\} \psi \left(\mathbf{r},t \right),
  \label{pauli_equation}
\end{equation}
where $\mathbf{\boldsymbol{\sigma}}=(\sigma_1,\sigma_2,\sigma_3)$ is a
vector synthesizing the Pauli matrices,
\begin{equation}
    \sigma_{1}=\begin{pmatrix}0 & 1\\
    1 & 0
\end{pmatrix},\;\sigma_{2}=\begin{pmatrix}0 & -i\\
    i & 0
\end{pmatrix},\;\sigma_{3}=\begin{pmatrix}1 & 0\\
    0 & -1
\end{pmatrix},\label{pauli_matrices}
\end{equation}
understood as operators associated with spin.
In this equation, $\psi \left(\mathbf{r},t \right)$ is a two-component
wave function \cite{GREINER1994} -- a mathematical entity called a
\emph{spinor} represented by a column matrix with two components.
Incorporating the spinor into the wave function is a direct consequence
of the binary nature of spin.
Consequently, the solutions of the Pauli equation involve two
differential equations, as opposed to one in the case of the
Schr\"{o}dinger equation.
In this work, we will not go into the formal aspects of the spinor,
which can be found in \cite{Bade1953}.

Employing the identity
\begin{equation}
  (\mathbf{\boldsymbol{\sigma}}\cdot\mathbf{a})
  (\mathbf{\boldsymbol{\sigma}}\cdot\mathbf{b})=
  \mathbf{a}\cdot\mathbf{b}+
  i\mathbf{\boldsymbol{\sigma}}\cdot(\mathbf{a}\times\mathbf{b}), \label{pauli_matrices_relation}
\end{equation}
into Eq. \eqref{pauli_equation}, we can rewrite the Pauli equation in a
way that will be more useful throughout the article
\begin{equation}
    i\hbar\frac{\partial}{\partial t}\psi \left(\mathbf{r},t
    \right)=\left\{
      \frac{1}{2m}\left[\left(\mathbf{p}-\frac{e}{c}\mathbf{A}\right)^{2}
        -\frac{e\hbar}{c}\mathbf{\boldsymbol{\sigma}}\cdot\mathbf{B}\right]
      +e\phi\right\} \psi \left(\mathbf{r},t \right),
    \label{pauli_equation_rewritten}
\end{equation}
where $\mathbf{B}$ is the magnetic
field, defined as $\mathbf{B}=\boldsymbol{\nabla} \times \mathbf{A}$.
In the context of our discussion, the significance of the Pauli equation
must be underscored. By incorporating spin, even if forcefully, into the
Schr\"{o}dinger wave function, his model became the most accurate
representation of an electron interacting with an electromagnetic field
under a non-relativistic regime.

Another physicist who worked extensively with spin was Charles Darwin,
who, in 1927, managed to extend Schr\"{o}dinger's theory to encompass
spin and, in this way, derive the anomalous Zeeman effect
\cite{DARWIN1927A,DARWIN1927B}.
Even so, both Pauli and Darwin failed to integrate Special Relativity
and Quantum Mechanics.
In his article \cite{PAULI1927}, Pauli acknowledged that his model was
provisional and that a definitive theory required invariance under
Lorentz transformation.

\section{Dirac equation}
\label{sec:dirac}

Dirac, dissatisfied with the KG equation, believed that a linear
relativistic wave equation could be developed in terms of the temporal
derivative.
This conviction arose from \emph{transformation theory}
\cite{PAIS1988}, a concept to which he and Jordan contributed,
aimed at generalizing both matrix and wave mechanics
\cite{JAMMER1989}.
Upon comparing the Schr\"{o}dinger equation with the KG equation, it is
apparent that the Schr\"{o}dinger equation fulfills this criterion,
whereas the KG equation does not.
To understanding  why this linearity is vital in the development
of quantum theory, see \cite{Wigner1959,Bargmann1964}.
Dirac expressed his strong preference for transformation theory with his
statement: ``The transformation theory had become my darling. I was not
interested in considering any theory that would not fit my darling''
\cite{Dirac1979} as cited in \cite{PAIS1988}.
Notably, since the Dirac equation is of utmost importance for this
article, the steps concerning it will be elaborated upon in greater
detail, with a deeper exploration of specific topics compared to the
earlier models.
The derivation presented here emphasizes Dirac's initial insight, while
a more didactic derivation can be found in \cite{THALLER1992} and more
explicitly detailed in \cite{SMIRMOV2016}.

Thus, while ``playing with the equations'' \cite{PAIS1988}, Dirac
obtained
\begin{equation}
  \mathbf{\mathbf{p}}^{2}\cdot\mathbbm{1}_{2}
  =(\boldsymbol{\sigma}\cdot\mathbf{p})^{2}.
  \label{product_vectorpauli_and_vectormomentum}
\end{equation}
We can rewrite Eq. \eqref{product_vectorpauli_and_vectormomentum} as
\begin{equation}
    \left(p_{1}^{2}+p_{2}^{2}+p_{3}^{2} \right)^{1/2}=\sigma_{1}p_{1}+\sigma_{2} p_{2}+\sigma_{3}p_{3},\label{product_vectorpauli_and_vectormomentum_expanded}
\end{equation}
Dirac claimed that both he and Pauli derived those matrices
independently \cite{PAIS1988}.

How can we generalize
Eq. \eqref{product_vectorpauli_and_vectormomentum_expanded} to four,
instead of three, components of the momentum?
This consideration was made with Special Relativity in mind, where a
vector quantity has three spatial and one temporal components.
It should be noted that, from now on, to refer generically to the
indices of the physical quantities with four components $0,1,2,3$, we
will use Greek letters, while we use Latin indices only for the spatial
or $1,2,3$ components.
This choice was made with relativistic covariant notation in mind to
avoid confusion in Section \ref{sec:relativistic_notation}.

In the case of momentum, the fourth component can be intuited by
dimensional analysis through Eq. \eqref{relativistic_energy}, so that
$p_{0}={E}/{c}$.
Dividing \eqref{relativistic_energy} by $c^{2}$ and incorporating the
time component, we have
\begin{equation}
  p_{0}^{2}-p_{1}^{2}-p_{2}^{2}-p_{3}^{2}
  =m^{2}c^{2}.
  \label{four_component_momentum_mass_c_squared}
\end{equation}
then, we take the square root and consider only the positive signal so
that
\begin{equation}
    (p_{0}^{2}-p_{1}^{2}-p_{2}^{2}-p_{3}^{2})^{1/2}=mc.
    \label{four_component_momentum_mass_c}
\end{equation}
It was desired to obtain, now with the four components of momentum, a
relation similar to
Eq. \eqref{product_vectorpauli_and_vectormomentum_expanded}.
Thus,
\begin{align}
    (p_{0}^{2}-p_{1}^{2}-p_{2}^{2}-p_{3}^{2})^{1/2} & =\gamma_{0}p_{0}-\gamma_{1}p_{1}-\gamma_{2}p_{2}-\gamma_{3}p_{3}, \label{four_component_momentum_gamma_matrices}
\end{align}
where Dirac conjectured the coefficients that multiplied the momentum
components.
We can compare  Eq. \eqref{four_component_momentum_mass_c} with Eq. \eqref{four_component_momentum_gamma_matrices} and reach the conclusion
\begin{equation}
  \gamma_{0}p_{0}-\gamma_{1}p_{1}-\gamma_{2}p_{2}-\gamma_{3}p_{3}=mc.
  \label{four_component_momentum_gamma_matrices_mass_c}
\end{equation}

Now, we must ask ourselves what these $\gamma$ coefficients are.
To find out, we square
Eq. \eqref{four_component_momentum_gamma_matrices_mass_c} and compare it
with Eq. \eqref{four_component_momentum_mass_c_squared}.
The following conditions are met for these coefficients,
\begin{align}
  \label{gamma_matrices_properties}
  (\gamma_{0})^{2} = {} & \mathbbm{1}_{4}, & \nonumber \\
  (\gamma_{i})^{2} = {} & -\mathbbm{1}_{4},& i = 1,2,3\\
  \gamma_{\mu}\gamma_{\nu} = {} & - \gamma_{\nu}\gamma_{\mu},
                                           &\mbox{for } \mu \neq\nu.
                                             \nonumber
\end{align}
The equality in the last line represents the anticommutation property of
the gamma matrices, $\{\gamma_{\mu},\gamma_{\nu}\}=0$ and the  nullity
of the anticommutator means that the coefficients are not numbers, since
for two nonzero numbers $z$ and $w$, $zw + wz \neq 0$.
Initially, Dirac believed that the Pauli matrices would fulfill this
role, since they obey the anticommutation relation.
However, he needed four matrices instead of three, which led to
\emph{gamma matrices}, whose minimal dimension is $4 \times 4$
(see Appendix \ref{sec:Appendix}).
Typically, these matrices are named \emph{Dirac gamma matrices}, and
$\gamma$ denotes them, yet we resisted using this notation to prevent
confusion with the Lorentz factor.
Any matrix that satisfies the algebra in Eq.
\eqref{gamma_matrices_properties} can be used;
however, the most usual representation for the study of the dynamics of
the model is the \emph{Dirac representation}, in which they are
\begin{align}
  \label{gamma_matrices_dirac_representation}
  \gamma_0 = {}
  &
    \begin{pmatrix}
       1 &  0 &  0 &  0\\
       0 &  1 &  0 &  0\\
       0 &  0 & -1 &  0\\
       0 &  0 &  0 & -1
    \end{pmatrix},
  &
    \gamma_1 = {}
  &
    \begin{pmatrix}
       0 &  0 &  0 &  1\\
       0 &  0 &  1 &  0\\
       0 & -1 &  0 &  0\\
      -1 &  0 &  0 &  0
    \end{pmatrix},
  \nonumber \\
  \gamma_2 = {}
  &
    \begin{pmatrix}
       0 &  0 &  0 & -i\\
       0 &  0 &  i &  0\\
       0 &  i &  0 &  0\\
      -i &  0 &  0 &  0
    \end{pmatrix},
  &
    \gamma_3 = {}
  &
    \begin{pmatrix}
       0 &  0 &  1 &  0\\
       0 &  0 &  0 & -1\\
      -1 &  0 &  0 &  0\\
       0 &  1 &  0 &  0
    \end{pmatrix},
\end{align}
which we can be simplified by using the Pauli's matrices
\begin{align}
  \gamma_{0} = {}
  &
    \begin{pmatrix}
      \mathbbm{1}_2 & 0\\
      0 & \mathbbm{-1}_2
    \end{pmatrix},
  &
    \gamma_{i} = {}
  &
    \begin{pmatrix}
      0 & \sigma_{i}\\
      -\sigma_{i} & 0
    \end{pmatrix},\label{eq:54}
\end{align}
with $i=1,2,3$.
We can rewrite Eq. \eqref{four_component_momentum_gamma_matrices_mass_c}
multiplying both sides by $c$ and manipulating it in such a way that
\begin{equation}
    c\gamma_{0}p_{0}=c(\gamma_{1}p_{1}+\gamma_{2}p_{2}+\gamma_{3}p_{3})+mc^{2},
\end{equation}
where we employ the definition of $p_{0}$, so that
\begin{equation}
    \gamma_{0}E=c(\gamma_{1}p_{1}+\gamma_{2}p_{2}+\gamma_{3}p_{3})+mc^{2},
    \label{gamma_energy_momentum_mass_c}
\end{equation}
which is an unusual way of writing the Dirac equation.
We can rewrite the Dirac equation making the definitions
\begin{equation}
  \label{gamma_as_functions_of_alpha_beta}
  \beta=\gamma_{0},\qquad
  \alpha_i=\beta\gamma_{i},\qquad
\end{equation}
where the $\beta$ and $\alpha$ matrices obey
\begin{equation} \label{anticommutator_alpha_beta}
    \{\alpha_i,\alpha_j\}=\{\alpha_i,\beta\}=0,
\end{equation}
and are called \emph{Dirac matrices}.
In the Dirac representation, we can write the $\alpha_i$ matrices simply
as
\begin{equation}
\alpha_{i}=\begin{pmatrix}0 & \mathbf{\mathbf{\mathbf{\sigma}}}_{i}\\
\mathbf{\sigma}_{i} & 0. \label{alpha_matrices_dirac_representation}
\end{pmatrix}.
\end{equation}
Multiplying \eqref{gamma_energy_momentum_mass_c} by $\beta$ from the
left and remembering the condition
$(\gamma_{0})^2=\beta^2=\mathbbm{1}_{4}$ from
\eqref{gamma_matrices_properties}, we obtain
\begin{equation}
    E=c(\alpha_{1}p_{1}+\alpha_{2}p_{2}+\alpha_{3}p_{3})+\beta mc^{2},
\end{equation}
which we can rewrite using the scalar product between
$\boldsymbol{\alpha}=(\alpha_1,\alpha_2,\alpha_3)$ and $\mathbf{p}$
as
\begin{equation}
    E=c\boldsymbol{\alpha}\cdot\mathbf{p} +\beta mc^{2},
\end{equation}
Using the operator in Eq. \eqref{energy_operator}, we have
\begin{equation}
  i\hbar\frac{\partial}{\partial t}
  =c\boldsymbol{\alpha}\cdot\mathbf{p}+\beta mc^{2}. \label{operators_of_dirac_equation}
\end{equation}
Operators, however, must act on a state.
Thus applying Eq. \eqref{operators_of_dirac_equation} on a state
$\psi \left(\mathbf{r},t \right)$ we obtain the \emph{Dirac equation}
\begin{equation}
  \label{dirac_equation}
  i\hbar\frac{\partial}{\partial t}\psi \left(\mathbf{r},t
  \right)=\left(c\boldsymbol{\alpha}\cdot\mathbf{p}
    +\beta mc^{2} \right) \psi \left(\mathbf{r},t \right).
\end{equation}

An intriguing story from this era, as detailed in \cite{Gottfried2011},
deserves attention.
When visiting Bohr's institute, the Danish physicist questioned Dirac
about his ongoing research. Dirac replied that he was trying to compute
the square root of a matrix, a claim that certainly puzzled Bohr.
What would captivate Bohr even more was learning—only after the
groundbreaking papers on the relativistic wave equation were
published—that Dirac was endeavoring to find the square root of the
\emph{identity matrix}.

Like the KG equation, there is an alternative way to express
Eq. \eqref{dirac_equation}, using covariant relativistic notation.
However, \eqref{dirac_equation} proves to be efficient when analyzing
the dynamics of the problems discussed here.
The additional noteworthy point is that $\psi(\mathbf{r},t)$, due to the
dimensions of $\boldsymbol{\alpha}$ and $\beta$, is a four-component
spinor referred to as a bispinor or Dirac spinor
\begin{equation}
  \psi(\mathbf{r},t)=
  \begin{pmatrix}
    \psi_{1}\\
    \psi_{2}\\
    \psi_{3}\\
    \psi_{4}
  \end{pmatrix},
  \label{dirac_spinor}
\end{equation}
where $\psi_i$ are the spinor components and we omitted variables in it
to simplify the notation.
It is crucial to note that the components of $\psi(\mathbf{r},t)$
represent wave functions but do not correspond to the four relativistic
dimensions.
Instead, they introduce new degrees of freedom for the particle, a
consequence of the simultaneous linearity in time and space.
The nature of these new components will be clarified by studying the
dynamics of free particles.

From comparison with Eq. \eqref{generalized_schrodinger_equation}, we
obtain the \emph{Dirac Hamiltonian}
\begin{equation}
  \label{dirac_hamiltonian}
  H=c\boldsymbol{\alpha}\cdot\mathbf{p}+\beta mc^{2},
\end{equation}
from which we conclude that, for $H$ to be Hermitian, the matrices
$\boldsymbol{\alpha}$ and $\beta$ must be Hermitian also, that is, they
are equal to their conjugate transposes:
$\boldsymbol{\alpha}=\boldsymbol{\alpha}^{\dagger}$ and
$\beta=\beta^{\dagger}$.

As we did with the previous wave equations, let us analyze the
conformity of the Dirac equation under the principles of  Relativity.
Replacing the momentum with its respective operator, we have
\begin{equation}
  \label{dirac_equation_laplacian}
  i\hbar\frac{\partial }{\partial t}\psi \left(\mathbf{r},t \right)=
  \left(-i\hbar c\boldsymbol{\alpha}\cdot\nabla+\beta mc{^2} \right)
  \psi (\mathbf{r},t).
\end{equation}
Thus, we observe that the temporal and spatial derivatives are both
linear and, therefore, have the same order.
Hence, stating that the Dirac equation is invariant under the Lorentz
transformation is reasonable.
For now, we will accept this argument as true and take for granted the
invariance of the Dirac equation.
However, in the next section, we will revisit this discussion and
provide a proof for this characteristic.

Just as we did with the KG equation, let us analyze the case of free
particles.
Firstly, we consider a particle at rest.
Although it is a restricted case, it will provide us with the necessary
insight regarding the allowed energy values.
We employ the Dirac equation with $\mathbf{p}=0$, so that
\begin{equation}
  i\hbar\frac{\partial}{\partial t}\Psi(\mathbf{r},t)
  =\beta mc^{2} \Psi \left(\mathbf{r},t \right),
  \label{free_particle_rest_dirac_equation}
\end{equation}
or in its matricial form
\begin{align}
  i\hbar\frac{\partial}{\partial t}
  \begin{pmatrix}
    \Psi_{1}\\ \Psi_{2}\\ \Psi_{3}\\  \Psi_{4}
  \end{pmatrix}  = {}
  &
    mc^{2}
    \begin{pmatrix}
      1 &  0 &  0 &  0\\
      0 &  1 &  0 &  0\\
      0 &  0 & -1 &  0\\
      0 &  0 &  0 & -1
    \end{pmatrix}
    \begin{pmatrix}
      \Psi_{1}\\ \Psi_{2}\\ \Psi_{3}\\ \Psi_{4}
    \end{pmatrix}.
\end{align}

The solutions for this set of differential equations is seen to be
\begin{align}
  \Psi_{1} = {}
  &
  \exp(-i\frac{mc^{2}}{\hbar}t)
  \begin{pmatrix}
    1\\ 0\\ 0\\ 0
  \end{pmatrix},
  &
  \Psi_{2} = {}
  &
  \exp(-i\frac{mc^{2}}{\hbar}t)
  \begin{pmatrix}
    0\\ 1\\ 0\\ 0
  \end{pmatrix},\nonumber \\
  \Psi_{3} = {}
  &
  \exp(i\frac{mc^{2}}{\hbar}t)
  \begin{pmatrix}
    0\\ 0\\ 1\\ 0
  \end{pmatrix},
  &
  \Psi_{4} = {}
  &
  \exp(i\frac{mc^{2}}{\hbar}t)
  \begin{pmatrix}
    0\\ 0\\ 0\\ 1
  \end{pmatrix},
\end{align}

where normalization factors are currently disregarded.

The first two expressions align with the anticipated outcomes for a
free particle scenario, as suggested by the negative sign in the
exponent.
However, as we will explore in the subsequent case, the two lower
components deviate from the expected behavior and are associated with
negative energy.

Now, let us examine the example of a particle confined to the $z$ axis,
possessing a momentum of magnitude $p$.
We expect a solution in the form
\begin{equation}
  \label{solutions_free_particle_dirac_equation}
  \Psi \left(\mathbf{r},t \right)=
  u\exp \left[- \frac{i}{\hbar} \left(\varepsilon t-pz \right) \right],
\end{equation}
where $u$ represents bispinors,
\begin{equation}
  u =
  \begin{pmatrix}
    u_{1}\\    u_{2}\\  u_{3}\\  u_{4}
  \end{pmatrix},
\end{equation}
and $\varepsilon$ the energy.
Thus, taking into account the dynamics only in $z$ direction the Dirac,
we have
\begin{equation}
    \left(\alpha_{3}pc+\beta mc^2 \right)u=\varepsilon u.
    \label{dirac_equation_free_particle_u_spinor}
\end{equation}
or equivalently using the matricial form of $\alpha_3$ and $\beta$,  we
obtain
\begin{equation} \label{matrix_product_coupled}
    \begin{pmatrix}mc^{2} & 0 & pc & 0\\
    0 & mc^{2} & 0 & -pc\\
    pc & 0 & -mc^{2} & 0\\
    0 & -pc & 0 & -mc^{2}
    \end{pmatrix}\begin{pmatrix}u_{1}\\
    u_{2}\\
    u_{3}\\
    u_{4}
    \end{pmatrix}=\varepsilon\begin{pmatrix}u_{1}\\
    u_{2}\\
    u_{3}\\
    u_{4}
\end{pmatrix}.
\end{equation}
The above equation can be decoupled into a two sets of two coupled
equations, that is,
\begin{align}
    \begin{pmatrix}mc^{2} & pc\\
    pc & -mc^{2}
    \end{pmatrix}\begin{pmatrix}u_{1}\\
    u_{3}
    \end{pmatrix} & =\varepsilon\begin{pmatrix}u_{1}\\
    u_{3}
    \end{pmatrix},\label{system_of_equations_1_dirac_equation_free_particle}\\
    \begin{pmatrix}mc^{2} & -pc\\
    -pc & -mc^{2}
    \end{pmatrix}\begin{pmatrix}u_{2}\\
    u_{4}
    \end{pmatrix} & =\varepsilon\begin{pmatrix}u_{2}\\
    u_{4}
    \end{pmatrix}.\label{system_of_equations_2_dirac_equation_free_particle}
\end{align}
We can isolate $u_3$ in the first pair of equations to obtain
\begin{equation}
    u_{3}=\frac{\varepsilon-mc^{2}}{pc}u_{1}, \label{solution_system_of_equations_dirac_free_particle}
\end{equation}
which we can reinserted into
Eq. \eqref{system_of_equations_1_dirac_equation_free_particle} to obtain
the free particle for energy,
\begin{equation}
    \varepsilon=\pm \left(m^{2}c^{4}+p^{2}c^{2} \right)^{1/2}.
\end{equation}
In this way, in a manner similar to the KG equation, we obtain
negative-energy eigenvalues.
This was initially demonstrated by Klein \cite{Klein1929A} in 1929 when
he found that the Dirac equation allows transitions from positive to
negative energy eigenvalues. Initially, there were efforts to eliminate
negative energy solutions \cite{Kragh1981}.
However, with each attempt proving unsuccessful, these results appeared
to be more inherent in the theory.
In fact, in November 1929, Klein and Nishina \cite{Klein1929B},
considering transitions to negative energies, derived the Compton
scattering formula.

\begin{figure}
    \centering
    \includegraphics[width=0.65\columnwidth]{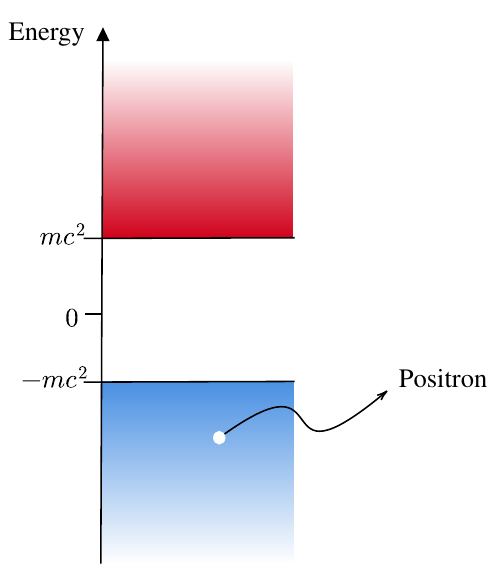}
    \caption{Dirac equation's energy diagram.
    }
    \label{fig:Energy Diagram}
\end{figure}

After some confusion \cite{PAIS1988}, Dirac managed to formulate a
coherent interpretation of this problem using the Pauli exclusion
principle \cite{DIRAC1930}.
He conjectured a sea of negative energy, referred to as
\emph{Dirac Sea}, in which all the negative-energy states are filled,
thus preventing an electron transition to negative energy.
According to Dirac's idea, a ``hole'' in this sea could behave like a
particle with positive charge.
Initially, Dirac speculated that the particle associated with the "hole"
was the proton.
This belief stemmed from the prevailing notion at the time that there
were only two elementary particles in nature.
However, as pointed out by Oppenheimer \cite{Oppenheimer1930}, the
presence of a proton in this sea of negative energy would lead to
annihilation and the release of energy in the form of photons, which
means, in a practical sense, the instability of matter.
Weyl \cite{Weyl1929}, employing the symmetry in the Maxwell and Dirac
equations, pointed out that the mass of the hole would have to be the
same as the electron.
In response to these critiques, Dirac stated, in an article published in
1931 \cite{Dirac1931}, that if there is a hole, it represents a new,
experimentally unknown particle with the same mass of the electron but
opposite charge -- an ``antielectron'', so to speak.
Today, this particle has been experimentally confirmed and is known as
the \emph{positron}.
Therefore, the new degree of freedom observed in the bispinor arises
precisely from the antiparticle associated with the particle.
The energy diagram in Fig. \ref{fig:Energy Diagram} illustrates the
concept of the Dirac Sea.

It is worth mentioning the Feynman-Stueckelber interpretation for
negative energies \cite{Feynman1948,Stueckelberg1942}, where negative
energy solutions are interpreted as positive energy particles moving
backward in time.
Reference \cite{GRIFFITHS2018} discusses the study of the free
particle with non-zero momentum, keeping this interpretation in mind.

To deepen our understanding of the meaning of negative energy, we can
build spinors for free particles.
Since the components of the Dirac spinor can be freely selected, we
focus on the configuration that provides the most straightforward
physical insights.
Thus, we represent the Dirac spinor as a two-component spinor
\begin{equation}
  \label{dirac_spinor_as_two_component_spinor}
  \Psi=
  \begin{pmatrix}\varphi \\
    \chi
  \end{pmatrix},
\end{equation}
where each component is a two-row matrix such that
\begin{equation}
  \varphi=
  \begin{pmatrix}
    \Psi_{1}\\
    \Psi_{2}
  \end{pmatrix},
  \qquad
  \chi=
  \begin{pmatrix}
    \Psi_{3}\\
    \Psi_{4}
  \end{pmatrix},
\end{equation}
and, similarly, for $u$,
\begin{equation}
  u_{A}=
  \begin{pmatrix}
    u_{1}\\
    u_{2}
  \end{pmatrix},
  \qquad
  u_{B}=
  \begin{pmatrix}
    u_{3}\\
    u_{4}
  \end{pmatrix},
\end{equation}
where $\varphi$ and $u_{A}$ correspond to the upper part of the Dirac
spinors and $\chi$ and $u_{B}$ correspond to the lower part.

For an energy $\varepsilon=+\varepsilon_{p}=\left(m^{2}c^{4}+p^{2}c^{2}
\right)^{1/2}$, we can make $u_{1}=1$ (and $u_{2}=u_{4}=0$). As a
consequence of this choice and according to Eq. \eqref{system_of_equations_1_dirac_equation_free_particle}, we obtain
\begin{equation}
  u_{3}=\frac{pc}{\varepsilon_{p}+mc^{2}},
\end{equation}
such that
\begin{equation}
  u_{A}=
  \begin{pmatrix}
    1\\
    0
  \end{pmatrix}
  \qquad
  u_{B}=
  \begin{pmatrix}
    \frac{pc}{\varepsilon_{p}+mc^{2}}\\
    0
  \end{pmatrix}.
\end{equation}
In the non-relativistic limit ($pc\ll\varepsilon_{p}+mc^{2}$), we have
the predominance of the upper component $\varphi$,  over the lower
component $\chi$.
If, still for $\varepsilon=+\varepsilon_{p}$, we impose $u_{2}=1$ (and $u_{1}=u_{3}=0$), so that
\begin{equation}
  u_{A}=
  \begin{pmatrix}
    0\\
    1
  \end{pmatrix},
  \qquad
  u_{B}=
  \begin{pmatrix}0\\
    -\frac{pc}{\varepsilon_{p}+mc^{2}}
  \end{pmatrix},
\end{equation}
once again, we observe that in the non-relativistic limit, there is a
predominance of the upper component of the spinor over the lower one.
In other words, for positive energies and in the non-relativistic
regime, the upper components of the Dirac spinor, as well as exhibits
dominance over the lower components.

On the other hand, if we take the negative energy
$\varepsilon=-\varepsilon_{p}$, we can set $u_{3}=1$ (hence
$u_{2}=u_{4}=0$), resulting in
\begin{equation}
  \label{u_A_and_u_B_1}
  u_{A}=
  \begin{pmatrix}
    -\frac{pc}{\varepsilon_{p}+mc^{2}}\\
    0
  \end{pmatrix}
  \qquad
  u_{B}=
  \begin{pmatrix}
    1\\
    0
  \end{pmatrix}.
\end{equation}
Furthermore, if we impose, still for $\varepsilon=-\varepsilon_{p}$, $u_{4}=1$ (consequently $u_{1}=u_{3}=0$), we find
\begin{equation}
  \label{u_A_and_u_B_2}
  u_{A}=
  \begin{pmatrix}
    0\\
    \frac{pc}{\varepsilon_{p}+mc^{2}}
  \end{pmatrix}
  \qquad
  u_{B}=
  \begin{pmatrix}
    0\\
    1
  \end{pmatrix}.
\end{equation}
Analyzing Eqs. \eqref{u_A_and_u_B_1} and \eqref{u_A_and_u_B_2}, we find
that in the non-relativistic limit, for negative energies, the lower
component $\chi$ of the Dirac spinor has predominance over the upper
component $\varphi$.

In other words, in the non-relativistic limiting case, it only requires
a degree of freedom in the spinor, just like in the Pauli equation.
That degree of freedom is associated with the spin of the particle in
the case of positive energy and with the spin of the antiparticle in the
case of negative energy.
However, in the case of the Dirac equation, the spin arises
spontaneously from the bispinor.

We will now study the manifestation of spin in the Dirac equation by
analyzing the helicity of the possible solutions in the free particle.
We can express the four bispinors that satisfy the Dirac equation and
arise from our arbitrary choices.
For positive energies, we have
\begin{equation}
  \label{u_R_and_u_L_1}
  u_{R}^{+} = {}
    N
    \begin{pmatrix}1\\
      0\\
      \frac{pc}{\varepsilon_{p}+mc^{2}}\\
      0
    \end{pmatrix},
    \qquad
    u_{L}^{+} =
    N
    \begin{pmatrix}
      0\\
      1\\
      0\\
      -\frac{pc}{\varepsilon_{p}+mc^{2}}
    \end{pmatrix},
\end{equation}
while for negative energies
\begin{equation}
  \label{u_R_and_u_L_2}
  u_{R}^{-} = {}
    N
    \begin{pmatrix}
      -\frac{pc}{\varepsilon_{p}+mc^{2}}\\
      0\\
      1\\
      0
    \end{pmatrix},
    \qquad
    u_{L}^{-}=
    N
    \begin{pmatrix}
      0\\
      \frac{pc}{\varepsilon_{p}+mc^{2}}\\
      0\\
      1
    \end{pmatrix},
\end{equation}
where the upper indices are linked to the sign of the energy and the
lower ones are associated with the helicity.
The calculation of the normalization factor $N$ will be performed subsequently, following the discussion on the probabilistic interpretation of the Dirac equation.

Inserting Eqs. \eqref{u_R_and_u_L_1} and \eqref{u_R_and_u_L_2} into
Eq. \eqref{solutions_free_particle_dirac_equation}, we can obtain the
solutions associated with each bispinor, which are given by
\begin{equation}
\begin{aligned}
 \Psi_{R}^{\pm}  &=  u_{R}^{\pm}\exp\left[\mp\frac{i}{\hbar}\left(\varepsilon_{p}t-pz
    \right)\right], \\
\Psi_{L}^{\pm}   &=  u_{L}^{\pm}\exp\left[\mp\frac{i}{\hbar}\left(\varepsilon_{p}t-pz
    \right)\right].
\end{aligned}
\label{solutions_free_particle_psi}
\end{equation}

One may ponder whether it is permissible to discard negative-energy
solutions, deeming them physically unacceptable.
The answer to that is in the negative, as a quantum system requires a
complete set of linearly independent states, and positive-energy
solutions alone do not suffice.
For the same reason, we cannot attempt to construct solutions solely
with positive-energy bispinors, i.e., assuming
$\varepsilon=+\left|\varepsilon \right|$, as this would result in
dependent bispinors.

In Eqs. \eqref{u_R_and_u_L_1} and \eqref{u_R_and_u_L_2}, the lower
subscripts $R$ and $L$ were not properly explained.
These subscripts are associate with the inherent helicity associated
with each bispinor.
To explore this characteristic, we must discuss the notion of operators
again.
In contemporary  Quantum Mechanics, operators can be represented by
matrices operating on states, represented by column matrices.
We can leverage this notation to facilitate our interpretation.

For the spin operator $\mathbf{S}$, it is expressed as
\begin{equation}
  \label{spin_operator}
  \mathbf{S}=\frac{\hbar}{2}\mathbf{\boldsymbol{\Sigma}},
\end{equation}
where
\begin{equation}
  \label{matrix_helicity}
  \mathbf{\boldsymbol{\Sigma}}
  \equiv
  \begin{pmatrix}
    \mathbf{\mathbf{\mathbf{\boldsymbol{\sigma}}}} & 0\\
    0 & \mathbf{\mathbf{\mathbf{\boldsymbol{\sigma}}}}
    \end{pmatrix}.
\end{equation}
When the operator $\mathbf{S}$ acts on a state, it performs a
measurement of spin, providing information concerning the spin of the
given state.
Similarly, when the Hamiltonian operator $H$ acts on a state, it
extracts information about the energy, for instance whether it is
positive or negative.
Hence, it becomes imperative to define a \emph{helicity operator},
which, when acting on a state, provides information about the helicity
of the corresponding state.
This operator is expressed as
\begin{equation}
  \Lambda =
  \mathbf{S} \cdot\frac{\mathbf{p}}{\left|\mathbf{p}\right|} =
  \frac{\hbar}{2}
  \left(
    \mathbf{\boldsymbol{\Sigma}}\cdot
    \frac{\mathbf{p}}{\left|\mathbf{p}
      \right|}
  \right),
\end{equation}
where the inner product represents the projection of spin onto the
momentum.
Therefore, when $\mathbf{S}$ and $\mathbf{p}$ are orthogonal, their
inner product is zero.

The behavior of $\Lambda$, in the case of free particle movement
restricted to $z$ is
\begin{equation}
  \Lambda=\mathbf{S}\cdot\frac{\mathbf{p}}{\left|\mathbf{p}\right|}
  =\frac{1}{\left|\mathbf{p}\right|}S_z
  =\frac{\hbar}{2\left|\mathbf{p}\right|}\Sigma{}_{z},
  \label{helicity_operator_z}
\end{equation}
that is, the projection of $\mathbf{S}$ onto $\mathbf{p}$ is
precisely the $S_z$ component of the spin. We have for $\Sigma{}_{z}$,
according to Eq. \eqref{matrix_helicity},
\begin{equation}
  \Sigma{}_{z}=
  \begin{pmatrix}
    1 & 0 & 0 & 0\\
    0 & -1 & 0 & 0\\
    0 & 0 & 1 & 0\\
    0 & 0 & 0 & -1
  \end{pmatrix},
\end{equation}
so that the helicity operator measures the helicity of the solutions by
the sign, positive or negative, of the elements along its diagonal when
applied to a given state.
For instance, for the state $u_L^{+}$, we find
\begin{align}
  \Lambda u_{L}^{+} = {}
  &
    \frac{\hbar}{2\left|\mathbf{p}\right|}N
    \begin{pmatrix}
      1 & 0 & 0 & 0\\
      0 & -1 & 0 & 0\\
      0 & 0 & 1 & 0\\
      0 & 0 & 0 & -1
    \end{pmatrix}
    \begin{pmatrix}
      0\\
      1\\
      0\\
      -\frac{pc}{\varepsilon_{p}+mc^{2}} \nonumber
    \end{pmatrix},\\
  = {}
  &
    -\frac{\hbar}{2\left|\mathbf{p}\right|}u_{L}^{+},
\end{align}
where the negative sign denotes negative helicity, commonly known as
left-handed helicity.
Applying the helicity operator to a positive, or right-handed, helicity
solution would result in the constant
(${\hbar}/{2\left|\mathbf{p}\right|}$) being multiplied by a
positive sign.
The distinct sign and its associated helicity confer the spin attribute
to the respective solution.
It is worth noting that we employ $u_{L}^{+}$ for simplicity in this
example, but the same helicity would hold for $\psi_{L}^{+}$ since what
holds significance is the sign acquired during the measurement process.
Thus, the exponential term in $\psi_{L}^{+}$ does not affect the outcome
of the measurement.

In summary, the action of the helicity operator provides information
regarding the connection between momentum and spin.
Positive helicity indicates parallel alignment between spin projection
and momentum, while negative helicity signifies anti-parallel alignment.
In the context of a free particle moving exclusively in the $z$
direction, the projected spin component is precisely the spin in the $z$
direction.
By convention, we denote that right-handed helicity corresponds to spin
up in this direction, whereas left-handed helicity corresponds to spin
down.

Applying the helicity operator, as represented by Eq.
\eqref{helicity_operator_z}, to the expressions given by
Eqs. \eqref{u_R_and_u_L_1} and \eqref{u_R_and_u_L_2}, leads to different
helicities for both positive and negative energy solutions.
For the positive energy solution $u_{R}^{+}$, the result is right-handed
helicity, while for $u_{L}^{+}$, it yields left-handed helicity, as
illustrated earlier.
In the context of the non-relativistic limit, where the upper components
have dominance, the coefficients $u_{1}$ and $u_{2}$ correspond to
\emph{spin up} and \emph{spin down}, respectively.
However, in the negative energy scenario, the application of
\eqref{helicity_operator_z} projects the right-handed helicity onto
$u_{R}^{-}$ and the left-handed helicity onto $u_{L}^{-}$.
In the non-relativistic limit, $u_{3}$ and $u_{4}$ correspond to
\emph{spin up} and \emph{spin down}, respectively.

\begin{figure}
  \label{fig:helicity}
  \centering
  \includegraphics[width=0.55\columnwidth]{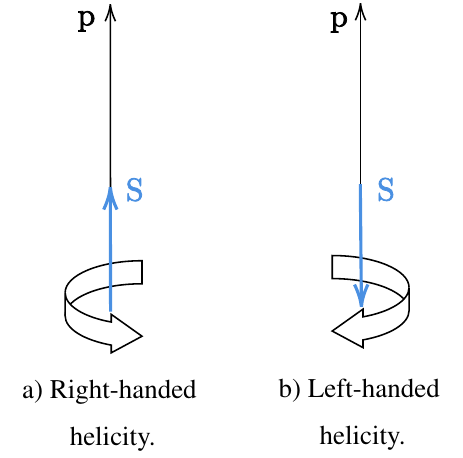}
  \caption{
    Illustrated depiction of helicity.
    In the left panel a) the alignment of momentum and spin in the same
    direction resulting in positive helicity, or right-handedness.
    Conversely, in the right panel b) the opposing directions of momentum
    and spin lead to negative helicity, or left-handedness.
  }
\end{figure}

Given that positive energy pertains to particles and negative energy
pertains to antiparticles, we may put aside brevity in favor of a more
explanatory notation of the solutions for a free particle.
In the following, we will adopt the notation found in
\eqref{solutions_free_particle_psi}, applying our findings collectively
to the states
\begin{equation}
    \begin{aligned}
    \Psi_{R}^{+} & \longrightarrow\Psi_{\rm up}^{\rm particle},\\
    \Psi_{L}^{+} & \longrightarrow\Psi_{\rm down}^{\rm particle},\\
    \Psi_{R}^{-} & \longrightarrow\Psi_{\rm up}^{\rm antiparticle},\\
    \Psi_{L}^{-} & \longrightarrow\Psi_{\rm down}^{\rm antiparticle}.
    \end{aligned}
\end{equation}

\begin{figure*}
  \centering
  \includegraphics[width=0.65\textwidth]{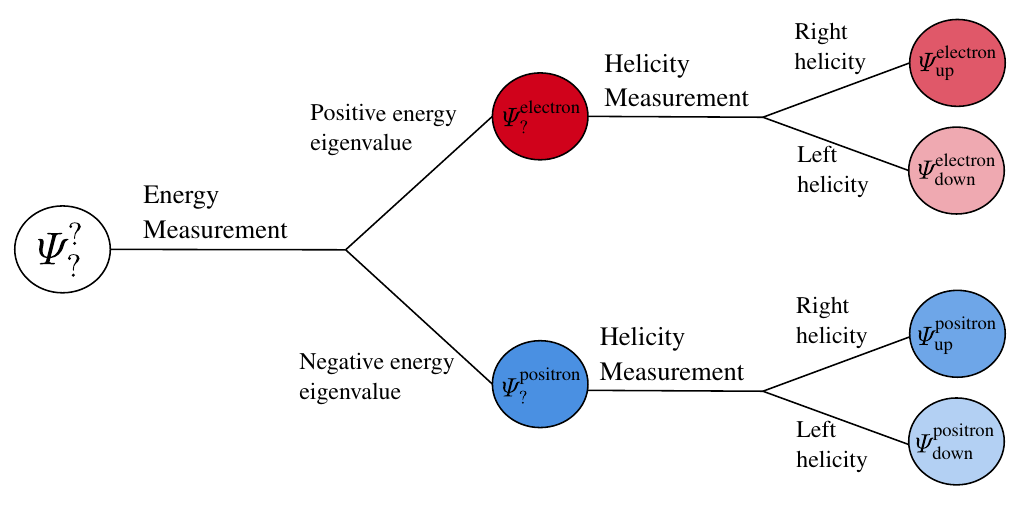}
  \caption{
    Illustration of the measurement energy and helicity of a given
    solution of the free particle situation.
    The measurement is performed on a state for which we initially do
    not know the sign of the energy or the type of helicity.
  }
  \label{fig:measuring_energy_and_helicity}
\end{figure*}

For didactic purposes, let us consider the following scenario:
we are in possession of a free particle solution $\Psi_{\rm ?}^{\rm ?}$,
but we do not know if it corresponds to a particle or antiparticle,
nor its spin.
However, we know that the state is associated with a electron or a
positron.
The process of discovering these two pieces of information is carried
out through measurements, with energy $H$ and helicity $\Lambda$
operators.
This process is illustrated in
Fig. \ref{fig:measuring_energy_and_helicity}.
Suggested as supplementary reading, the reference \cite{Neamtam1952}
provides an explicit derivation of solutions for the case of a free
particle in three dimensions.

Having covered the energy of the free particle, the next step involves
testing the continuity equation to determine whether the density
$\rho(\mathbf{r},t)$ can be defined as positive.
We initiate this process by multiplying \eqref{dirac_equation_laplacian}
by the conjugate transpose of $\psi$,
$\psi^{\dagger}=\left(\psi_{1}^{*},\psi_{2}^{*},\psi_
  {3}^{*},\psi_{4}^{*}\right)$,
obtaining
\begin{equation}
  i\hbar\psi^{\dagger}\frac{\partial\psi}{\partial t}
  =-ic\hbar\psi^{\dagger}\boldsymbol{\alpha}\cdot\mathbf{\nabla\psi}
  +mc^2\psi^{\dagger}\beta\psi,
  \label{multiplying_dirac_equation_by_psi_dagger}
\end{equation}
that we take the conjugate transpose as
\begin{equation}
  -i\hbar\psi\frac{\partial\psi^{\dagger}}{\partial t} =
  ic\hbar\psi\boldsymbol{\alpha}\cdot\mathbf{\nabla\psi^{\dagger}}
  +mc^2\psi\beta\psi^{\dagger}.
  \label{multiplying_dirac_equation_by_psi_dagger_dagger}
\end{equation}

Subtracting Eq. \eqref{multiplying_dirac_equation_by_psi_dagger_dagger}
from Eq. \eqref{multiplying_dirac_equation_by_psi_dagger}, we have
\begin{equation}
  \psi^{\dagger}\frac{\partial\psi}{\partial t}
  +\psi\frac{\partial\psi^{\dagger}}{\partial t}=-
  c\psi^{\dagger}\boldsymbol{\alpha}\cdot\mathbf{\nabla\psi}
  -c\psi\boldsymbol{\alpha}\cdot\mathbf{\nabla\psi^{\dagger}},
  \label{subtraction_psi_psi_dagger}
\end{equation}
where we used the Hermitian property of $\boldsymbol{\alpha}$ and
$\beta$.
Employing the product rule, we rewrite
Eq. \eqref{subtraction_psi_psi_dagger} as
\begin{equation}
    \frac{\partial\left|\psi\right|^{2}}{\partial t}+\nabla(c\psi^{\dagger}\boldsymbol{\alpha}\psi)=0, \label{continuity_equation_dirac_equation}
\end{equation}

Comparing the above equation with the continuity equation we conclude
that
\begin{equation} \rho(\mathbf{r},t)=\left|\psi\right|^{2}=
  \left|\psi_{1}\right|^{2}+\left|\psi_{2}\right|^{2}
    +\left|\psi_{3}\right|^{2}+\left|\psi_{4}\right|^{2},
    \label{rho_density_dirac_equation}
\end{equation}
where $\rho(\mathbf{r},t)$ is clearly positive definite since it is the
sum of the squares of the magnitudes of the components of $\psi
\left(\mathbf{r},t \right)$, which allow us to have an probability
interpretation for $\rho(\mathbf{r},t)$.
Furthermore, the probability current is defined as
\begin{equation}    \mathbf{j}(\mathbf{r},t)=
  c\psi^{\dagger}\boldsymbol{\alpha}\psi.
  \label{j_current_dirac_equation}
\end{equation}

It is common to use the notation
$\bar{\psi}=\psi^{\dagger}\beta=\psi^{\dagger}\gamma^{0}$ resulting in
\begin{align}
    \rho(\mathbf{r},t) & =\bar{\psi}\gamma_{0}\psi,\\
    \mathbf{j}(\mathbf{r},t) & =c\bar{\psi}\mathbf{\mathbf{\gamma}}\psi.
\end{align}

We can interpret Eq. \eqref{rho_density_dirac_equation} as the
probability of finding a given particle in a certain region.
If we consider a single particle restricted to a volume $V$, we rewrite
it as
\begin{equation}
    \rho(\mathbf{r},t)=\frac{1}{V}, \label{probability_density_volume}
\end{equation}
that we can use to calculate the normalization factor $N$ of the free
particle solutions.
Taking the bispinor in Eq. \eqref{solutions_free_particle_psi}
associated with $u_{R}^{+}$ as an example, we compute
\begin{align}
  \psi_{R}^{+\dagger}\psi_{R}^{+} = {}
  & \frac{1}{V},\nonumber \\
  u_{R}^{+\dagger}u_{R}^{+} = {}
  & \frac{1}{V},\nonumber \\
  N^{2}\left[1+\frac{p^{2}c^{2}}{\left(\varepsilon_p
  +mc^{2}\right)^{2}}\right]
  = {}
  &
   \frac{1}{V},
\end{align}
where we isolate $N$ to find the expression
\begin{equation}
    N=\sqrt{\frac{\varepsilon_p+mc^{2}}{2\varepsilon_p V}},
\end{equation}
which is the same normalization factor as the other solutions to the
free particle problem. 

We add that the modern attitude toward the Dirac equation differs from
the one presented here, where we interpret it in a manner similar to the
Schrödinger equation.
This discrepancy arises because the ``single-particle'' interpretation
breaks down, as the creation and annihilation of particles become
possible in nontrivial interactions. 
The proper framework to address this issue is quantum field theory (QFT) \cite{SAKURAI2020,THIBES2022}.

The elucidation of plane wave solutions and their admissible negative
energies, along with a viable probabilistic interpretation, underscores
the strong aspects of Dirac's model.
However, the question remains:
Does the Dirac equation derive the Pauli equation in the
non-relativistic scenario?

\subsection{Electromagnetic interactions in the non-relativistic limit}
\label{subsec:eletro_nr}

In this subsection, we will examine how the Dirac equation behaves in
the presence of electromagnetic interactions within the non-relativistic
limit, precisely with the aim of derive the Pauli equation.
We start by incorporating the minimal coupling into Dirac Hamiltonian,
as given in Eq. \eqref{minimal_coupling_substituition}.
This results in
\begin{equation}
    H=c\boldsymbol{\alpha}\cdot\boldsymbol{\Pi}+\beta mc^{2}+e\phi. \label{dirac_hamiltonian_minimal_coupling}
\end{equation}
Thus, applying $H$ on $\Psi \left(\mathbf{r},t \right)$, Eq. \eqref{dirac_spinor_as_two_component_spinor}, we obtain
\begin{equation}
  \left(c\boldsymbol{\alpha}\cdot\boldsymbol{\Pi}
    +\beta mc^{2}+e\phi \right)\Psi \left(\mathbf{r},t \right)=
  \varepsilon\Psi \left(\mathbf{r},t \right),
\end{equation}
where $\varepsilon$ is an energy eigenvalue described as
\begin{equation}
    \varepsilon=K+e\phi+mc^{2},
\end{equation}
where $K$ is the kinetic energy, $e\phi$ is the electric potential
energy and $mc^{2}$ is the rest energy.
We can write this eigenvalue equation in its matrix form as
\begin{equation}
    \begin{pmatrix}mc^{2}+e\phi & c\boldsymbol{\sigma}\cdot\boldsymbol{\Pi}\\
    c\boldsymbol{\sigma}\cdot\boldsymbol{\Pi} & -mc^{2}+e\phi
    \end{pmatrix}\begin{pmatrix}\varphi\\
    \chi
    \end{pmatrix}=\varepsilon\begin{pmatrix}\varphi\\
    \chi
    \end{pmatrix}, \label{system_of_equations_eletromagnetic_dirac}
\end{equation}
from which we can extract a system of two equations as follows.
From the lower equation, we have
\begin{equation}
    c\left(\boldsymbol{\sigma}\cdot \mathbf{\boldsymbol{\Pi} } \right)\varphi=\left(\varepsilon+mc^{2}-e\phi \right)\chi,
\end{equation}
which can be approximated to
\begin{equation}
    \left(\boldsymbol{\sigma}\cdot \mathbf{\boldsymbol{\Pi} } \right)\varphi\approx2mc^{2}\chi, \label{approximation_dirac_eletromagnetic}
\end{equation}
since, for positive energies, the kinetic and electric energy are much
smaller than $mc^{2}$ in the non-relativistic limit.
Similarly, we can write the upper equation as
\begin{equation}
    c\left(\boldsymbol{\sigma}\cdot \mathbf{\boldsymbol{\Pi} } \right)\chi+e\phi\varphi=(\varepsilon_{NR})\varphi, \label{solution_system_eletromagnetic_dirac}
\end{equation}
where $\varepsilon_{NR}=K+e\phi$, is the non-relativistic energy, that
is, the total energy excluding the rest energy.
Replacing  Eq. \eqref{approximation_dirac_eletromagnetic} into
Eq. \eqref{solution_system_eletromagnetic_dirac}, we obtain
\begin{equation}
  \frac{(\boldsymbol{\sigma}\cdot\boldsymbol{\Pi})(\boldsymbol{\sigma}
    \cdot\boldsymbol{\Pi})}{2m}\varphi+e\phi\varphi=\varepsilon_{NR}\varphi.
\end{equation}
Using the identity related to the the Pauli matrices products, we obtain
\begin{equation}
  \left[\frac{\boldsymbol{\Pi}^{2}}{2m}
    +\frac{i\boldsymbol{\sigma}}{2m}\cdot
    \left(\boldsymbol{\Pi}\times\boldsymbol{\Pi} \right)
    +e\phi \right]\varphi=\varepsilon_{ NR}\varphi,
  \label{numerator_solution_system_eletromagnetic_dirac}
\end{equation}
where we can rewrite the vector product
$(\boldsymbol{\Pi}\times\boldsymbol{\Pi})$ by applying it to  $\varphi$,
i.e,
\begin{align}
  \left(\boldsymbol{\Pi}\times\boldsymbol{\Pi} \right)\varphi
  = {}
  &
    \left(\mathbf{p}-\frac{e}{c}\mathbf{A} \right)\times
    \left(\mathbf{p}-\frac{e}{c}\mathbf{A} \right)\varphi,
    \nonumber \\
  = {}
  &
    \left(-i\hbar\boldsymbol{\nabla}-\frac{e}{c}\mathbf{A}\right)\times
    \left(-i\hbar\boldsymbol{\nabla}\varphi-\frac{e}{c}\mathbf{A}\varphi
    \right),
    \nonumber \\
  = {}
  &
    i\hbar\frac{e}{c}\left[\boldsymbol{\nabla}\times
    \left(\mathbf{A}\varphi \right)+\mathbf{A}\times
    \left(\boldsymbol{\nabla}\varphi \right) \right],\nonumber \\
  ={}
  &
    i\hbar\frac{e}{c}\left(\boldsymbol{\nabla}\times
    \mathbf{A} \right)\varphi, \nonumber \\
  = {}
  &
    i\hbar\frac{e}{c}\mathbf{B}\,\varphi,
\end{align}
so that,  Eq. \eqref{numerator_solution_system_eletromagnetic_dirac}
becomes
\begin{equation}
    \left(\frac{\mathbf{\boldsymbol{\Pi}}^{2}}{2m}-\frac{\hbar e}{2mc}\boldsymbol{\sigma}\cdot\mathbf{B}+e\phi \right)\varphi=\varepsilon_{NR}\varphi,
\end{equation}
which is, precisely, the Pauli equation [see
Eq. \eqref{pauli_equation_rewritten}]
In this manner,  it can be stated that the Pauli equation represents the
non-relativistic limit of the Dirac equation, and the emergence of the
Pauli matrices, linked to the spin, is a natural outcome of Dirac's
theory.
In particular, the term $(e \hbar/2mc)\boldsymbol{\sigma}$ represents
the magnetic moment of the electron, incorporating the gyromagnetic
ratio factor of $g=2$, which had previously been assumed \emph{ad hoc}.

\subsection{Foldy-Wouthuysen transformation}
\label{subsec:fw_trans}

Another, more sophisticated, and general method for testing the
non-relativistic limit of the Dirac equation exists.
This approach is known as the Foldy-Wouthuysen (FW) transformation
\cite{Foldy1950}, introduced by Leslie Foldy and Siegfried Wouthuysen
(and later generalized by Caze \cite{Caze1954}), and we will delve into
it in this subsection.
Costella and McKellen argue that it is only through this method that a
significant classical limit is achieved with respect to particles and
antiparticles \cite{Costella1995}.
We have chosen to highlight this procedure because of its historical
importance and the broad range of relevant applications.
For instance, deriving the Dirac equation in a rotating frame must
reduce to the Pauli equation in a rotating frame in the non-relativistic
limit \cite{Matsuo2011}.
Additionally, we can connect the final result obtained with the FW
transformation into the Dirac Hamiltonian with the Sommerfeld equation
[see Eq. \eqref{sommerfeld_formula}] and the fine structure of hydrogen.

However, to accomplish this, we utilize a more demanding mathematical
framework.
Therefore, it is natural for a reader who has not encountered
difficulties thus far to encounter them now.
Furthermore, we will carry out this procedure in the case of interaction
with an electromagnetic field.
We will not delve into every step of the calculations due to their
length, but this derivation can be found in more detail in the
refs. \cite{GREINER2000,Bjorken1964}.

As we saw in the free particle solution, the positive (negative) energy
solutions have a large (small) components.
The main concept of the FW transformation is to utilize this
characteristic of the Dirac equation as a foundation for a new
representation. For this purpose, the concept of parity is used.

An operator or matrix is even (odd) if it commutes (anti-commutes) with
the parity matrix.
In practice, an odd operator connects elements of opposite parity, while
an even operator connects elements of similar parity.
In the relativistic scenario, we can consider the example of the
Dirac matrix $\alpha_i$, which ``connects'' - in this case, couples -
the elements of positive and negative energy of the bispinor.
On the other hand, the matrix $\beta$ is even, since it does not couple
the positive and negative energy elements.
Thus, we can rewrite the Dirac Hamiltonian as
\begin{equation}
  \label{eq:odd_even_Dirac_Hamiltonian_minimal_coupling}
  H=\beta mc^{2}+\mathcal{O}+\mathcal{E},
\end{equation}
where $\mathcal{O}$ is the odd part of the Dirac Hamiltonian and
$\mathcal{E}$ is the even part.
Thus,
\begin{align}
  \label{eq:odd_even}
  \mathcal{O} = {} & c\boldsymbol{\alpha}\cdot\mathbf{\Pi},\nonumber\\
  \mathcal{E} = {} & e\phi.
\end{align}
Due to the parity property, we have the following relations:
\begin{align} \label{eq:parity_beta_relations}
    \beta\,\mathcal{O} = {} &-\mathcal{O}\,\beta, \nonumber\\
    \beta\,\mathcal{E} = {} & \mathcal{E}\,\beta.
\end{align}
Previously, we noted that our goal is to eliminate the odd component
from the Hamiltonian, since it links the upper and lower sections of the
bispinor.
To achieve this, we apply a transformation in the Hamiltonian,
ensuring that in the resulting form the odd component is no longer
present.
To do this, we define the operator
\begin{equation}
    S=-\frac{\mathrm{i}}{2mc^{2}}\beta\,\mathcal{O},
\end{equation}
and during the derivation we perform unitary transformations of the type
\begin{equation}
  \label{eq:unitary_transformation_fw_dirac_hamiltonian}
    H^{\prime}=\exp(iS)H\exp(-iS),
\end{equation}
where we have assumed that the field, and consequently, the vector
potential and the Hamiltonian are time-independent.
This unitary transformation is a FW transformation.
The above equation should be understood in the light of a power series,
using the Baker-Hausdorff lemma
\begin{align}
  e^{iL}Ae^{-iL} = {}
  &
    [A+i\left[L,A\right]+\frac{(i)^{2}}{2!}
    \left[L,\left[L,A\right]\right]+\notag\\
  &
    \ldots+\frac{(i)^{n}}{n!}\left[L \left[L,\left[L,\ldots,L\left[L,A\right],\ldots\right]\right]\right]+\ldots\,,
\end{align}
where $\left[L,A\right] = LA - AL$ is the commutator between the
operators $L$ and $A$.
Therefore, Eq. \eqref{eq:unitary_transformation_fw_dirac_hamiltonian}
becomes
\begin{align}
  H^{^{\prime}}= {}
  &
    H+
    i\left[S,H \right]
    +\frac{(i)^{2}}{2!}\left[S,\left[S,H \right]\right]
    \nonumber \\
  &
    +\frac{(i)^{3}}{3!}\left[S,\left[S \left[S,H \right] \right] \right]
    +\frac{(i)^{4}}{4!}\left[S,\left[S,\left[S\left[S,H\right]\right]\right]
    \right]
    \ldots \,.
    \label{eq:expanded_transformed_hamiltonian}
\end{align}
when we consider terms up to the order of $1/m^3c^6$.
We calculate the commutators separately, up to the desired order
employing Eq. \eqref{eq:parity_beta_relations} as follows
\begin{equation}
  i\left[S,H \right]=-
  \mathcal{O}+\frac{1}{2mc^{2}}\,\beta\left[\mathcal{O},\mathcal{E}\right]
  +\frac{1}{mc^{2}}\beta\,\mathcal{O}^{2},
\end{equation}
\begin{align}
  \frac{i^{2}}{2!}\left[S,\left[S,H \right] \right] = {}
  &
    -\frac{1}{2mc^{2}}\,\beta\,\mathcal{O}^{2}-\frac{1}{2m^{2}c^{4}}\,\mathcal{O}^{3} \notag\\
  & -\frac{1}{8m^{2}c^{4}}\left[\mathcal{O},\left[\mathcal{O},\mathcal{E}\right]\right],
\end{align}
\begin{align} \label{eq:third_commutator}
  \frac{i^{3}}{3!}\left[S,\left[S \left[S,H \right] \right] \right] = {}
  &
    \frac{1}{6m^{2}c^{4}}\mathcal{O}^{3}-\frac{1}{6m^{3}c^{6}}\,\beta\,\mathcal{O}^{4}
    \notag \\
  & -\frac{1}{48m^{3}c^{6}}\,\beta\left[\mathcal{O},\left[\mathcal{O},\left[\mathcal{O},\mathcal{E}\right]\right]\right],
\end{align}
\begin{equation}
    \frac{(i)^{4}}{4!}\left[S,\left[S,\left[S \left[S,H \right] \right] \right] \right] =\frac{1}{24m^{3}c^{6}}\beta\,\mathcal{O}{}^{4}.
\end{equation}
For the odd part, it suffices to include terms up to the order of
$1/m^{2}c^{4}$, and hence the third term in Eq. \eqref{eq:third_commutator}
can be neglected.
In this way, we can rewrite
Eq. \eqref{eq:expanded_transformed_hamiltonian} as
\begin{align}
  H^{\prime} ={}
  & \beta
    \left(mc^{2}+\frac{\mathcal{O}^{2}}{2mc^{2}}
    -\frac{\mathcal{O}^{4}}{8m^{3}c^{6}}\right)
    +\mathcal{E}
    -\frac{1}{8m^{2}c^{4}}
    \left[\mathcal{O},\left[\mathcal{O},\mathcal{E}\right]\right]\notag\\
  &
    + \frac{1}{2mc^{2}}\beta\left[\mathcal{O},\mathcal{E}\right]
    -\frac{1}{2mc^{2}}\,\mathcal{O}^{3}.\label{prime_Dirac_Hamiltonian}
\end{align}
Now, considering that $\mathcal{O}$ raised to even powers and
$\left[\mathcal{O},\left[\mathcal{O},\mathcal{E}\right]\right]$ are
even, there are still odd terms in the equation.
Therefore, we can rewrite it as
\begin{equation}
    H^{\prime}=\beta mc^{2}+\mathcal{E}^{^{\prime}}+\mathcal{O}^{^{\prime}}.
\end{equation}
Thus, we write
\begin{align}
  \mathcal{O}^{^{\prime}} = {}
  & \frac{1}{2mc^{2}}\beta\left[\mathcal{O},\mathcal{E}\right]
    -\frac{1}{3m^{2}c^{4}}\mathcal{O}^{3},\\
  \mathcal{E}^{^{\prime}} = {}
  & \mathcal{E}+\beta\left(\frac{\mathcal{O}^{2}}{2mc^{2}}
    -\frac{\mathcal{O}^{4}}{8m^{3}c^{6}}\right)
    -\frac{1}{8m^{2}c^{4}}
    \left[\mathcal{O},\left[\mathcal{O},\mathcal{E}\right]\right].
    \label{el}
\end{align}
The next step consists in applying the operator
\begin{equation} \label{eq:prime_s}
    S^{\prime}=-\frac{\mathrm{i}}{2mc^{2}}\beta\,\mathcal{O^{\prime}},
\end{equation}
so that the Hamiltonian in Eq. \eqref{prime_Dirac_Hamiltonian}, after
undergoing a second FW transformation, becomes
\begin{equation} \label{double_prime_dirac_hamiltonian}
  H^{\prime\prime}=\beta mc^{2}+\mathcal{E}^{^{\prime}}
  +\frac{i}{2mc^{2}}\beta[\mathcal{O}^{^{\prime}},\mathcal{E}^{^{\prime}}].
\end{equation}
The third term in the above equation is odd, so it is necessary to
perform a third FW transformation so that we obtain
\begin{equation}
    H^{\prime\prime\prime}=\beta mc^{2}+\mathcal{E}^{\prime}\equiv
    H_{\Phi},
    \label{Hp}
\end{equation}
where $H_{\Phi}$ is given by
\begin{equation}
    \!\!H_{\Phi}=\beta mc^{2}+\beta\frac{\mathcal{O}^{2}}{2mc^{2}}-\beta\frac{\mathcal{O}^{4}}{8m^{3}c^{6}}+e\phi-\frac{1}{8m^{2}c^{4}}\left[\mathcal{O},\left[\mathcal{O},\mathcal{E}\right]\right].
\end{equation}
Using the relation
$(\boldsymbol{\alpha}\cdot\mathbf{a})(\boldsymbol{\alpha}\cdot\mathbf{b})=\mathbf{a}\cdot\mathbf{b}+i\,\boldsymbol{\Sigma}\cdot(\mathbf{a}\times\mathbf{b})$
and $\mathbf{E}=-\nabla\phi$, with calculations similar to those applied
in the previous subsection, we obtain
\begin{align}
  H_{\Phi} = {}
  & \beta
    mc^{2}+\beta\frac{1}{2m}\left(\mathbf{p}-\frac{e}{c}\mathbf{A}\right)^{2}-\frac{e\mathrm{\hbar}}{2mc}\beta\left(\boldsymbol{\Sigma}\cdot\mathbf{B}\right)-\frac{\beta\,\mathbf{p}^{4}}{8m^{3}c^{6}}+e\phi
    \nonumber \\
  &
    -\frac{e\hbar^{2}}{8m^{2}c^{2}}\boldsymbol{\nabla}\cdot\mathbf{E}
    -\frac{ie\hbar^{2}}{8m^{2}c^{2}}
                   \boldsymbol{\Sigma}\cdot(\boldsymbol{\nabla}\times\mathbf{E})
    -\frac{e\hbar }{4m^{2}c^{2}}
    \boldsymbol{\Sigma}\cdot(\mathbf{E}\times\mathbf{p}).
\end{align}
Taking the case of positive energy, we can replace $\beta$ and $\boldsymbol{\Sigma}$ with their upper $2\times2$ blocks
\begin{equation}
    \begin{aligned}
    \beta & \rightarrow\mathbb{\mathbbm{1}}_2,\\
    \boldsymbol{\Sigma} & \rightarrow\boldsymbol{\sigma}.
    \end{aligned}
\end{equation}
Thus, we obtain
\begin{align}
  \label{eq:Hamiltonian_Phi_representation}
  H_{\Phi} = {}
  &
    mc^{2}+\frac{1}{2m}\left(\mathbf{p}
    -\frac{e}{c}\mathbf{A}\right)^{2}
    -\frac{\mathbf{p}^{4}}{8m^{3}c^{6}}
    -\frac{e\mathrm{\hbar}}{2mc}\boldsymbol{\sigma}\cdot\mathbf{B}
    +e\phi \nonumber \\
  &
    -\frac{ie\hbar^{2}}{8m^{2}c^{2}}\boldsymbol{\sigma}\cdot
    \left(\boldsymbol{\nabla}\times\mathbf{E}\right)
    -\frac{e\hbar}{4m^{2}c^{2}}\boldsymbol{\sigma}
    \cdot\left(\mathbf{E}\times\mathbf{p}\right)
    \nonumber\\
    &
    -\frac{e\hbar^{2}}{8m^{2}c^{2}}
    \boldsymbol{\nabla}\cdot\mathbf{E}.
\end{align}
The first three terms above represent the expansion of the kinetic
energy
$K=[\left(\mathbf{p}-e/c\mathbf{A}\right)^{2}+mc^{2}]^{1/2}-mc^{2}$,
describing the increase in mass in the relativistic case.
The next pair of terms is associated with magnetic dipole and
electrostatic energy.
Subsequently, the next two terms describe the spin-orbit (SO)
interaction.
For a spherically symmetric potential, we have
$\nabla\times\mathbf{E}=0$.
Additionally, we can write
\begin{equation}
  \boldsymbol{\sigma}\cdot(\boldsymbol{E}\times\mathbf{p})
  =-\frac{1}{r}\frac{\partial\phi}{\partial
    r}\boldsymbol{\sigma}\cdot(\mathbf{r}\times\mathbf{p})
  =-\frac{1}{r}\frac{\partial\phi}{\partial r}(\boldsymbol{\sigma}\cdot\mathbf{L}),
\end{equation}
so that the SO can be written as
\begin{equation}
  H_{\rm SO}=\frac{\hbar e}{4m^{2}c^{2}}
  \frac{1}{r}\frac{\partial\phi}{\partial r}
  (\boldsymbol{\sigma}\cdot\mathbf{L}).
\end{equation}
Finally, The last term in Eq. \eqref{eq:Hamiltonian_Phi_representation}
is called \emph{Darwin term}, and it is associated with a purely
relativistic effect, the \emph{Zitterbewegung effect}, which translates
to "trembling motion of the electron" from German, as coined by
Schr\"{o}dinger \cite{SCHRöDINGER1930}.
This steams from the interference between  the positive- and
negative-energy eigenstates.

In this new representation, the Hamiltonian $H_{\Phi}$ contains terms
from the Pauli equation, added to the relativistic terms and the
SO interaction.
The presence of these latter terms revisits an aspect mentioned earlier:
the fine structure of hydrogen.

There are many other interesting phenomena related to the Dirac equation.
In this context, we cite two topics that highlight novel aspects of the
equation and serve as extensions of the themes discussed here.
Interested readers may further explore them by treating these topics as
exercises, with guidance from appropriate textbooks
\cite{Bjorken1964,GREINER1994}.
These include:
\begin{enumerate}
\item
  Study the \emph{Zitterbewegung} effect by computing the Heisenberg
  equation of motion for the position operator.
\item
  Study the Klein paradox, by deriving the transmission and reflection
  coefficients in the scattering of an electron hitting a potential step
  \cite{Klein1929A}.
\end{enumerate}

As mentioned previously, Schr\"{o}dinger encountered difficulties in
deriving Sommerfeld's formula through his relativistic approach.
If the Dirac equation indeed proves to be the suitable equation for
describing quantum and relativistic phenomena, then
Eq. \eqref{sommerfeld_formula} should be deducible from it.
Initially, Dirac utilized a first-order approximation method akin to
Pauli's earlier approach.
However, in 1928, Darwin \cite{Darwin1928} and Gordon \cite{Gordon1928}
successfully obtained an exact solution, showcasing that the
experimentally successful result of old quantum theory could be derived
from a more formal representation.

Dirac published his theory in two papers, \cite{DIRAC1928A,DIRAC1928B},
the first of which was published in February.
The effect of Dirac's theory of the electron was revolutionary in
Quantum Mechanics.
Rosenfeld would call it a ``miracle, an absolute marvel''
\cite{Kragh1981}.
From one of the biggest initial obstacles of the Dirac equation, the
allowed negative energies came one of its greatest triumphs: the
prediction of antiparticles.
In 1931, as mentioned, Dirac conjectured an antiparticle to fill the sea
of negativity, but there was no experimental proof.
In 1933, however, Anderson \cite{ANDERSON1933} experimentally proved the
existence of positrons, solidifying the Dirac equation as the most
accurate equation for representation of spin 1/2 particles, a notion
that endures in Relativistic Quantum Mechanics to this day.

\section{Relativistic covariant notation}
\label{sec:relativistic_notation}

This section presents natural units and relativistic covariant notation
as alternative approaches to express the KG and Dirac equations.
Utilizing these methods will enable a more straightforward
classification of how equations transform under Dirac transformations.
Even though natural units and relativistic notation can be used
separately, we opt to use them together here to improve the practical
utility of the equations discussed.
As mentioned in the Introduction, we aim to enhance the experience of
exploring the KG and Dirac equations through these notations. 

In natural units, physical constants serve as the units of associated
physical quantities.
Although this approach may appear arbitrary, its implications become
evident in the expressions of various physical quantities.
The speed of light $c=3.0\times10^{8} \mbox{ }m/s$, for example, can be
used as a natural unit of speed, so that a speed of $v=1.0\times10^{8}
\mbox{ } m/s$ becomes $v=1.0\times10^{8} \mbox{ } m/sc$, or $1/3$,
simply.
Therefore, $v$ becomes a dimensionless parameter, which we usually
designate $\beta$.
Time, elementarily defined as $distance/v$, is measured in
\emph{distance units}.
For the purposes of this work, we will use, in addition to $c=1$, the
normalized Planck constant $\hbar=1$.
Using this we can write the KG \eqref{kg_equation_lhs} and the Dirac
equation \eqref{dirac_equation_laplacian} respectively as 
\begin{align}
    \left(\frac{\partial^{2}}{\partial
  t^{2}}-\boldsymbol{\nabla}^{2}+m^{2}\right)\psi \left(\mathbf{r},t
  \right)= {} & 0,\label{kg_equation_natural_units} \\
    \left(-i\boldsymbol{\alpha}\cdot\boldsymbol{\nabla}+\beta m \right)
  \psi \left(\mathbf{r},t \right)= {} & i\frac{\partial}{\partial t}\psi \left(\mathbf{r},t \right). \label{dirac_equation_natural_units}
\end{align}

As mentioned earlier, in  Relativity, we deal with four
dimensions: one temporal and three spatial, which can be interpreted as
four components of a vector. Physical quantities composed of these four
components are termed four vectors.
We implicitly utilize this concept when discussing momentum, even
though, at that point, we were \emph{not} employing the covariant
notation.
There are two types of four-vectors: contravariant and covariant; this
classification stems from differential geometry and will not be explored
in this work.
A generic contravariant four-vector is denoted by an upper index, being
expressed as 
\begin{equation}
  a^{\mu} =\left(a^{0},a^{1},a^{2},a^{3} \right)
  =\left(a^{0},\boldsymbol{a} \right),
\end{equation}
where $\boldsymbol{a}$ represents the three spatial components. We can write, following the most common notation,
\begin{equation}
    a^{\mu}=\left(a^{0},a^{i} \right),
\end{equation}
where the Greek indices ($\mu,\upsilon,\gamma$) run through $0,1,2,3$
and the Latin indices ($i,j,k$) go through $1,2,3$.
A contravariant four-vector $a^{\mu}$ has a dual covariant four-vector,
symbolized with the lower index $a_{\upsilon}$ 
\begin{equation}
    a_{\upsilon}=\eta_{\upsilon\mu}a^{\mu},\label{covariant_contravariant_relation}
\end{equation}
where $\eta_{\upsilon\mu}$ is the metric tensor in Minkowski space,
suitable for Special Relativity, and expressed in covariant
components as
\begin{equation}
\eta_{\upsilon\mu}=\begin{pmatrix}
    1 & 0 & 0 & 0\\
    0 & -1 & 0 & 0\\
    0 & 0 & -1 & 0\\
    0 & 0 & 0 & -1
    \end{pmatrix}. \label{matrix_tensor_minkowski_space}
\end{equation}

In Eq. \eqref{covariant_contravariant_relation}, we used Einstein sum
notation,
\begin{equation}
    \sum_{\mu,\nu=0}^{3}b_{\upsilon}c^{\mu}\rightarrow b_{\upsilon}c^{\mu}=b_{0}c^{0}+b_{1}c^{1}+b_{2}c^{2}+b_{3}c^{3},
\end{equation}
where the repeated index symbolizes summation.
Making the summation explicit and incorporating the chosen metric,
$a_\nu$ is expressed as
\begin{equation}
    a_{\nu} =(a^{0},-a^{1},-a^{2},-a^{3})=(a^{0},-\boldsymbol{a}).
\end{equation}

A relevant operation with four-vectors is the inner product, exclusively
applicable between a contravariant and a covariant four-vectors.
Consider a generic covariant four-vector
$b_{\mu}=(b^{0},-\boldsymbol{b})$.
The inner product between $b_{\mu}$ and $a^{\mu}$ is defined as
\begin{align}
  a^{\mu}b_{\mu} = {}
  &
    a^{0}b^{0}-a^{1}b^{1}-a^{2}b^{2}-a^{3 }b^{3},\nonumber \\
  = {}
  &
    a^{0}b^{0}-\boldsymbol{a}\cdot\boldsymbol{b},
\end{align}
or, more specifically, in the case of $b_{\mu}=a_{\mu}$ as
\begin{align}
  a^{\mu}a_{\mu} = {}
  &
  a^{0}a^{0}-a^{1}a^{1}-a^{2}a{}^{2}-a^ {3}a^{3},\nonumber \\
  = {}
  &
    \left(a^{0}\right)^{2}-\boldsymbol{a}^{2}.
\end{align}

Having defined the basic properties of the relativistic covariant notation, we can employ it to define physical quantities as four-vectors. The contravariant four-vector position is
\begin{align}
    x^{\mu} = {} & \left(x^{0},x^{1},x^{2},x^{3}\right),\nonumber \\
    x^{\mu} = {} & \left(x^{0},\boldsymbol{x} \right),\\
    x^{\mu} = {} & \left(t,\boldsymbol{x} \right),\nonumber
\end{align}
with the inner product $x_{\mu}x^{\mu}$ expressed as
\begin{equation}
    x_{\mu}x^{\mu}=\left(x^{0}\right)^{2}-\boldsymbol{x}^{2}.
\end{equation}

The definition of the position contravariant four-vector allows us to write the four-gradient operator
\begin{align}
  \frac{\partial}{\partial x^{\mu}} = {}
  &
    \left(\frac{\partial}{\partial x^{0}},\frac{\partial}{\partial x^{1}} ,\frac{\partial}{\partial x^{2}},\frac{\partial}{\partial x^{3}}\right),\nonumber \\
  = {}
  &
    \left(\partial_0,\partial_1,\partial_2,\partial_3 \right),\nonumber \\
  = {}
  &
    (\partial_t,\boldsymbol{\nabla}),\nonumber \\
  ={}
  &
    \partial_{\mu}, \label{four_gradient_operator}
\end{align}
which is covariant, even with the positive sign in the spatial
coordinates.
The contravariant dual is then defined as
\begin{equation}
    \partial^{\mu}=\left(\partial_0,-\boldsymbol{\nabla} \right),
\end{equation}
with the inner product $\partial_{\mu}\partial^{\mu}$ written as
\begin{equation}
    \partial_{\mu}\partial^{\mu}=\partial_{0}^{2}-\boldsymbol{\nabla}^{2}. \label{dalembertian_relativistic_notation}
\end{equation}

We can also define a four-moment, in covariant form as
\begin{align}
    p_{\mu} ={} & \left(p^{0},-p^{1},-p^{2},-p^{3} \right),\nonumber \\
    p_{\mu} ={} & \left(p^{0},-\mathbf{p}\right),\\
    p_{\mu} ={} & \left(E,-\mathbf{p}\right),\nonumber
\end{align}
with the inner product $p_{\mu}p^{\mu}$ expressed as
\begin{align}
    p_{\mu}p^{\mu} = {} &\left(p^{0}\right)^{2}-\mathbf{p}^{2}, \nonumber\\
    = {} & E^{2}-\mathbf{p}^{2}, \nonumber\\
    = {} & m^{2},
\end{align}
associated with the energy of a relativistic particle \eqref{relativistic_energy}.

Now, let us employ the techniques outlined to rewrite and study the KG and Dirac equations from an alternative perspective. By substituting \eqref{dalembertian_relativistic_notation} into \eqref{kg_equation_natural_units}, we can express the KG equation as
\begin{equation}
    \left(\partial_{\mu}\partial^{\mu}+m{{}^2} \right)\psi \left(\mathbf{r},t \right)=0. \label{kg_equation_relativistic_notation}
\end{equation}

In addition to providing an alternative way to write the equations, the relativistic covariant notation facilitates the analysis of invariance under Lorentz transformations. This happens because the equations expressed in this notation have the property of transforming predictably when the aforementioned transformations are applied.  Instead of conducting a proper Lorentz transformation, which would involve introducing additional concepts, we will confine our analysis to a mathematical argument grounded in the space-time interval, whose definition will be introduced shortly.

When assessing the invariance of the Dirac equation under Lorentz transformations, we invoked the symmetry between time and space dictated by Relativity, as evidenced by the derivative orders. This rationale is reinforced by the non-invariance of the Schr\"{o}dinger equation (with time as a first-order derivative and space as a second-order derivative) and the invariance of the KG equation (featuring second-order derivatives in both time and space). An alternative route to the same conclusion involves considering the space-time interval $ds^2$, defined as
\begin{equation}
    ds^2  = dx^{\mu}dx_{\mu} = dt^{2}-d\boldsymbol{x}^{2},
    \label{space_time_interval}
\end{equation}
must be invariant \cite{Landau2013}.  Consequently, it can be inferred that the combination of derivatives, expressed in covariant notation as $\partial_{\mu}\partial^{\mu}$, remains constant when transitioning between inertial frames of reference.

Upon examination of equation \eqref{kg_equation_relativistic_notation}, the presence of the inner product between the derivatives \eqref{dalembertian_relativistic_notation} and of the mass $m$ becomes evident. The mass, being the rest mass, remains constant and does not alter with a change of reference frame. Furthermore, as we have just established, the inner product is invariant. Consequently, utilizing covariant relativistic notation allows for an immediate recognition of the invariance of the KG equation under Lorentz transformations.

Multiplying \eqref{dirac_equation_natural_units} by $\beta$ by the left and rewriting it with covariant notation results in
\begin{equation}
    \left[i \left(\gamma^{0}\partial_{0}+\boldsymbol{\gamma}\cdot\boldsymbol{\nabla} \right)-m \right]\psi \left(\mathbf{r},t \right),
\end{equation}
where we identify an inner product between the four-vector $\gamma^{\mu}$ and the derivative $\partial_\mu$ \eqref{four_gradient_operator}, so that
\begin{equation}
    \left(i\gamma^{\mu}\partial_{\mu}-m \right)\psi \left(\mathbf{r},t \right),
\end{equation}
where we again identify the mass and an inner product. The product in question is not directly related to the space-time interval \eqref{space_time_interval}, and its invariance cannot be assumed immediately. It is imperative to demonstrate that the inner product within the Dirac equation remains invariant. To establish this, we will extend our proof to encompass any pair of contravariant and covariant four-vectors, aiming to generalize the result for both the KG and Dirac equations. To commence, we initiate a one-dimensional Lorentz transformation, akin to equations \eqref{x_another_frame} and \eqref{t_another_frame}, on two arbitrary four-vectors.
\begin{align*}
    a'^{0} ={} & \gamma \left(a^{0}-\beta a^{1} \right),\\
    a'^{1} ={} & \gamma \left(a^{1}-\beta a^{0} \right),\\
    b'_{0} ={} & \gamma \left(b_{0}-\beta b_{1} \right),\\
    b'_{1} ={} & \gamma \left(b_{1}-\beta b_{0} \right),
    \end{align*}
where $\beta=v$ in natural units. The other components remain the same, so the inner product after the Lorentz transformation is
\begin{align}
  a'^{\mu}b'_{\mu} = {}
  &
    a'^{0}b'^{0}-a'^{1}b'^{1}-a^{2}b^{2}-a^{3}b^{3} \nonumber \\
  = {}
  &
    \gamma \left(a^{0}-\beta a^{1} \right)\gamma \left(b_{0}-\beta b_{1} \right)\nonumber \\
    & -\left[\gamma \left(a^{1}-\beta a^{0} \right)\gamma(b_{1}-\beta b_{0})+a^{2}b^{2}+a^{3}b^{3} \right]\nonumber \\
  = {} &
         \gamma^{2} \left(1-\beta^{2} \right)a^{0}b_{0}-\gamma^{2} \left(1-\beta^{2} \right)a^{1}b_{1}-a^{2}b^{2}-a^{3}b^{3},
\end{align}
where we use \eqref{lorentz_factor_gamma}, which leads to
\begin{equation}
    a'^{\mu}b'_{\mu} = a^{0}b_{0}-a^{1}b_{1}-a^{2}b^{2}-a^{3}b^{3} = a^{\mu}b_{\mu}. \nonumber
\end{equation}
Here, we observe that the inner product between any two four-vectors
remains constant, consequently ensuring the constancy of the Dirac
equation.
This observation also extends to the KG equation. 

Furthermore, if one wishes to write the Dirac equation even more
bluntly, Feynman notation, where
$\gamma^{\mu}\partial_{\mu}=\slashed{\partial}$, is employed, so that 
\begin{equation}
    \left(i\slashed{\partial}-m \right)\psi \left(\mathbf{r},t \right)=0.
\end{equation}

To conclude the comparison between the equations, we present Table
\ref{tab:1}, aiming to encapsulate the most relevant characteristics of
the models presented in this work. 
\begin{table*}
    \centering
    \begin{adjustbox}{width=\textwidth,center}
    \begin{tabular}{ c c c c }
    \hline
    \hline
    \textbf{Feature} &
    \textbf{Schr\"{o}dinger} &
    \textbf{Klein-Gordon} &
    \textbf{Dirac}\\
    \hline
\textbf{Equation}
    &
    $
    i\hbar\frac{\partial }{\partial t}\psi =
    \left(-\frac{\hbar^2}{2m}\boldsymbol{\nabla}^{2} +V\right) \psi
    $
    &
    $
    \frac{1}{c^{2}}\frac{\partial ^{2}  }{\partial t^{2}}\psi
    =\left(\boldsymbol{\nabla }^{2}
    -\left(\frac{mc}{\hbar}\right)^{2}\right) \psi
    $
    &
    $
    i\hbar \frac{\partial}{\partial t}\psi =
    \left(-ic\hbar \boldsymbol{\alpha} \cdot
    \boldsymbol{\nabla} +\beta mc^{2}\right)\psi
    $
    \\
    \textbf{Energy}
    &
    Only positive eigenvalues.
    &
    Both positive and negative energies are allowed.
    &
    Both positive and negative energies are allowed.
    \\
    \textbf{Probabilistic interpretation}   &
    Viable, since $\rho >0$.
    &
    Not viable since $\rho$ is not positive definite.
    &
    Viable, since $\rho >0$.
    \\
    \textbf{Relativity}
    &
    Not-invariant under Lorentz transformation.
    &
    Invariant under Lorentz transformation.
    &
    Invariant under Lorentz transformation.
    \\
    \textbf{Spin}
    &
    Does not contain spin.
    &
    Does not contain spin.
    &
    Contains the spin.
    \\
    \hline
    \hline
    \end{tabular}
    \end{adjustbox}
    \caption{
      Table comparing the Schr\"{o}dinger, KG, and Dirac equations
      regarding the aspects addressed in this article.
    }
    \label{tab:1}
\end{table*}

\section{Summary}
\label{sec:conc}

In this work, we provided an introduction to the Dirac equation,
placing particular emphasis on the historical context of its conception.
In this context, we conducted comparisons with the Schr\"{o}dinger and
KG equations, focusing on their invariance under Lorentz transformations
and energy eigenvalues.

We started with a historical preamble, illustrating that, in the 1920s:
i) Physics was in a period of rapid development, marked by
new discoveries in Quantum Mechanics,
ii) a consensus on a consistent methodology had not yet been reached,
and
iii) Dirac was deeply involved in both previous aspects, actively
contributing to the construction of the new physics.
Moreover, we outlined features of matrix mechanics and underscored the
significance of Dirac's canonical quantization, which was found to be
applicable in other instances in this study.

The natural course was to present wave mechanics as an alternative to
matrix mechanics.
Schr\"{o}dinger's equation enabled many physicists \cite{BELLER1983} to
have a more practical view of  Quantum Mechanics and, from it, Born was
able to extract a probabilistic interpretation that originated from the
density $\rho(\mathbf{r},t)$ being positive definite.
Furthermore, Schr\"{o}dinger managed to solve the problem of the
eigenvalues for the hydrogen atom, deriving the non-relativistic hydrogen
spectra -- a feat that matrix mechanics had not been able to accomplish
without \emph{ad hoc} hypotheses.
However, the Schr\"{o}dinger equation is not invariant under the Lorentz
transformation; it does not fit into Relativity.
We show its non-invariance explicitly, with a one-dimensional Lorentz
transformation and without relativistic notation.

Subsequently, we introduced the KG equation.
Initially conceived as a wave equation compatible with Relativity
principles, it fulfills this purpose by remaining invariant under
Lorentz transformations.
However, upon delving into the dynamics of the model for a free
particle, we demonstrated that it permits solutions with negative
energies.
Moreover, the second-order time derivative impeded defining a density
$\rho(\mathbf{r},t)$ as positive, consequently preventing a
probabilistic interpretation.
Consequently, its physical applications were limited then, especially
considering the lack of knowledge about spinless particles. The
consideration of spin prompts us to focus on the subsequent wave
equation: the Pauli equation.

Spin, the electron's new degree of freedom, emerged to explain the
quantization detected in SG's experiment.
Pauli forcibly inserted spin into Schr\"{o}dinger's equation, replacing
the wave function with a spinor -- a two-component column matrix -- and
including his matrices.
Pauli successfully derived a precise equation applicable to the
non-relativistic regime through this approach.
Nevertheless, Pauli was unable to make progress regarding the
integration of Quantum Mechanics and Relativity.

Finally, we introduced the Dirac equation, marking the concluding
chapter of the narrative we aimed to tell.
Dirac, dissatisfied with the KG equation as the relativistic wave
equation due to the second-order time derivative that hindered a
probabilistic interpretation, sought a more fitting solution in line
with his transformation theory.
We presented a deduction of the equation from an elegant result found by
Dirac, discussing the emergence of the $4 \times 4$ matrices and the
bispinor as evidence of yet another degree of freedom beyond the spin.
It is noteworthy how matrices, seldom employed in Physics until the
advent of matrix mechanics, arise naturally as inherent to the Dirac
equation.

We encountered negative energy solutions once again after solving the
dynamics for a free particle at rest.
The deeper investigation into the physical significance of these
energies unfolds when considering the case of a moving particle, where
we have observed that the two upper components of the bispinor
correspond to the particle, associated with positive energy, while the
lower components correspond to an entity linked with negative energy.
We commented on how Dirac's conjecture, his sea of negative energy,
explains this result and postulates, through a ``hole'' in the sea, the
existence of an antiparticle.
Helicity serves as an auxiliary means of interpreting these outcomes:
Positive (right-handed) and negative (left-handed) helicity are imposed,
respectively, on the first and second components of the upper part of
the bispinor in the case of positive energy.
For negative energy, the third component of the bispinor exhibits
right-hand helicity, while the fourth component displays left-hand
helicity.
We have used these types of helicity to discern the spin of the
solutions that emerge from the free particle problem, just as we
employed the sign of energy to differentiate between the particle and
the antiparticle.

By analyzing the continuity equation for Dirac's equation, we establish
that the density $\rho(\mathbf{r},t)$ is positive definite, allowing for
a probabilistic interpretation.
This characteristic, at the time, underscored the primacy of the Dirac
equation over the KG equation.
Additionally, we demonstrated that, in the non-relativistic limit, the
Dirac equation yields the Pauli equation, naturally yielding the
magnetic spin ratio.
Furthermore, we commented on how the model manages to derive the
Sommerfeld formula exactly.
In this context, we have shown that terms associated with relativistic
corrections and spin-orbit interaction appear by employing the FW
transformation.
Finally, we use relativistic covariant notation to show more concise
forms of the KG and Dirac equations and, utilizing four-vectors, the
invariance of Dirac's model.

We recommend further reading of \cite{THIBES2022}, which informatively
elucidates how the initial incongruity between the Schr\"{o}dinger
equation and Relativity gives rise to Quantum Field Theory
(QFT) and underscores the pivotal role played by the Dirac equation in
facilitating this transition.

\section*{Acknowledgments}
This work was partially supported by the Brazilian agencies Conselho
Nacional de Desenvolvimento Científico e Tecnológico (CNPq) and
Instituto Nacional de Ciência e Tecnologia de Informação Quântica
(INCT-IQ).
It was also financed by the Coordenação de Aperfeiçoamento de Pessoal de
Nível Superior (CAPES, Finance Code 001).
F.M.A. acknowledges CNPq Grant No. 314594/2020-5.
E.O.S. acknowledges CNPq Grant 306308/2022-3, FAPEMA Grants APP-12256/22 and UNIVERSAL-06395/22.

\appendix
\section{The Dimension of the Dirac matrices}
\label{sec:Appendix}

In this section, we discuss the dimension of Dirac matrices, utilizing the \emph{bra(c)ket} notation introduced by Paul Dirac. Here, a \emph{ket} $|\zeta\rangle$ represents a state and can be expressed as a vector within a specific Hilbert space. Every ket has a corresponding bra $\langle \zeta|$, representing the same state but originating from a Hilbert space dual to the ket's space. Multiplying a ket by its corresponding bra yields
\begin{equation}
    \left\langle \zeta|\zeta\right\rangle =1 . \label{bra_ket_inner_product}
\end{equation}

We start the discussion on the dimension of Dirac matrices from the assumption that we do not know any representation for $\beta$ and $\alpha_i$, only the algebra of the anticommutators \eqref{anticommutator_alpha_beta} and
\begin{align}
    \beta^{2} =1, \label{eq:A.3}
\end{align}
a relation that can be obtained by recalling the definition $\gamma_0$ \eqref{gamma_as_functions_of_alpha_beta} and that $\gamma_0^2=1$\eqref{gamma_matrices_properties}. From the squaring of $\gamma_i$ as functions of $\alpha_i$, as indicated in \eqref{gamma_as_functions_of_alpha_beta}, we can similarly infer a relationship for $\alpha_i$, leading to
\begin{align}
    \gamma_{i}^2 &=(\beta\alpha_{i})^2, \nonumber \\
    & = \beta\alpha_{i}\beta\alpha_{i}. \label{gamma_squared_alpha_beta}
\end{align}
Through their anticommutator \eqref{anticommutator_alpha_beta}, we can write $\beta\alpha_{i}=-\alpha_{i}\beta$, so that \eqref{gamma_squared_alpha_beta} becomes
\begin{equation}
    \gamma_{i}^2=-\alpha_{i}\beta\beta\alpha_{i}.
\end{equation}
Moreover, we have $\gamma_i^2 = -1$ \eqref{gamma_matrices_properties} and \eqref{eq:A.3}, so that
\begin{equation}
    \alpha_i^2=1. \label{alpha_i_squared}
\end{equation}
Furthermore, we take $\alpha_{i}$ and $\beta$ as Hermitian for being part of the Hamiltonian $H$ \eqref{dirac_hamiltonian}.

To advance the discussion, we must analyze the diagonal elements of the matrices, a task we will undertake by employing the trace. The trace $\tr(A)$ of a matrix $A$
\begin{equation}
    A=\begin{pmatrix}a_{11} & \cdots & a_{1n}\\
    \vdots & \ddots & \vdots\\
    a_{n1} & \cdots & a_{nn}
    \end{pmatrix},
\end{equation}
is given by the sum of the elements on its diagonal
\begin{equation}
    \tr(A)=\sum_{i=1}^{n}a_{ii}=a_{11}+(...)+a_{nn}.
\end{equation}

Interestingly, the trace does not change if we multiply the matrix $A$ by a unitary matrix $U$ on the right and by the respective conjugate transpose unitary matrix $U^{\dagger}$ on the left
\begin{equation}
    \tr(A)=\tr(U^{\dagger}AU).
    \label{eq:A.5}
\end{equation}

But, as they are Hermitian, therefore $\alpha_{i}=\alpha_{i}^\dagger$ and $\beta=\beta^\dagger$, and we have the condition \eqref{eq:A.3}, the matrices $\alpha_{i}$ and $\beta$ are considered unitary, such that
\begin{align}
    \alpha_{i}^{\dagger}\alpha_{i} & =1,\\ \label{alpha_beta_unity}
    \beta^{\dagger}\beta & =1.
\end{align}

Using \eqref{eq:A.5} for $\alpha_{i}$, multiplied the other component of $\boldsymbol{\alpha}$ and its conjugate transpose, $\alpha_{j}$ and $\alpha_{j}^{\dagger}$, such that
\begin{equation}
    \tr(\alpha_{i})=\tr(\alpha_{j}^{\dagger}\alpha_{i}\alpha_{j}),
\end{equation}
where we can use the anticommutation relation \eqref{anticommutator_alpha_beta}, concluding that $\alpha_{i}\alpha_{j}=-\alpha_{j}\alpha_{i}$, and that the trace is
\begin{equation}
    \tr(\alpha_{i})=-\tr(\alpha_{j}^{\dagger}\alpha_{j}\alpha_{i}).
\end{equation}

Nevertheless, the product between the components in $j$ is equal to the unit \eqref{alpha_i_squared} such that
\begin{equation}
    \tr(\alpha_{i})=-\tr(\alpha_{i}),
\end{equation}
which is only satisfied if the trace of $\alpha_{i}$ is null
\begin{equation}
    \tr(\alpha_{i})=0.
\end{equation}
Repeating the same process for the $\beta$ matrix, we obtain
\begin{equation}
    \tr(\beta)=0.
\end{equation}

Now, let us analyze the action of $\alpha_{i}$ on a state $|\Lambda\rangle$, which is an eigenstate of $\alpha_{i}$ and, consequently, has the following eigenvalue relationship
\begin{equation}
    \alpha_{i}|\Lambda\rangle=\lambda|\Lambda\rangle,
\end{equation}
where $\lambda$ is the eigenvalue associated with $|\Lambda\rangle$.

Making the inner product between two of these eigenstates, as in \eqref{bra_ket_inner_product}, we find
\begin{equation}
    \left\langle \Lambda|\Lambda\right\rangle =1,
\end{equation}
where we can insert $\alpha_{i}^{\dagger}\alpha_{i}$, since they are equal to unity \eqref{alpha_i_squared}, applying them individually on the states, obtaining
\begin{align}
    \left\langle \Lambda|\Lambda\right\rangle & =\left\langle \Lambda\right|\alpha_{i}^{\dagger}\alpha_{i}\left|\Lambda\right\rangle,\nonumber \\
    & =\left\langle \Lambda\right|\lambda^{2}\left|\Lambda\right\rangle,
\end{align}
from where we conclude that
\begin{equation}
                \lambda^{2}=1\rightarrow\lambda=\pm1,
\end{equation}
given that the eigenvalues of a Hermitian matrix are real. Another way to calculate the trace of a given matrix is precisely through its eigenvalues
\begin{equation}
    \tr(\alpha_{i})=\sum_{l}\lambda_{l},
\end{equation}
where we use the index $l$ to avoid confusion with the index $i$ from the matrix $\alpha_{i}$. However, since the trace of $\alpha_{i}$ is zero, the number of elements on the main diagonal and the dimension of the matrix must be even, and the same can be said for $\beta$. For a $2 \times 2$ dimension, the three Pauli matrices satisfy the anti-commutation relation \eqref{anticommutator_alpha_beta}. However, since Dirac needed \emph{four} matrices, the minimum dimension for $\beta$ and $\alpha_i$ is $4 \times 4$. Naturally, according to \eqref{gamma_as_functions_of_alpha_beta}, the dimension of the gamma matrices is also $4\times4$.

\bibliographystyle{apsrev4-2}
%

\end{document}